\documentclass[journal]{IEEEtran}
\usepackage[cmex10]{amsmath}
\usepackage[noadjust]{cite}

\usepackage{url}
\usepackage{color}

\usepackage{amssymb}
\usepackage{path}
\usepackage{verbatim}
\usepackage{algorithm}
\usepackage{framed}
\usepackage{algpseudocode}

\usepackage{color}
\usepackage{amsmath}
\usepackage{cases}
\usepackage{theorem}
\usepackage[export]{adjustbox}

\usepackage{graphicx}

\usepackage{pgfplots}
\usepackage{pgfplotstable}
\pgfplotsset{compat=newest}

{ \theorembodyfont{\normalfont} 

}

\newcounter{enumctr}

\DeclareFontFamily{U}{mathx}{\hyphenchar\font45}
\DeclareFontShape{U}{mathx}{m}{n}{<-> mathx10}{}
\DeclareSymbolFont{mathx}{U}{mathx}{m}{n}
\DeclareMathAccent{\widebar}{0}{mathx}{"73}

\begin{document}
%
\title{Pedestrian-Aware Engine Management Strategies for Plug-in Hybrid Electric Vehicles}

\author{Yingqi Gu, 
	Mingming Liu,
	Joe Naoum-Sawaya,
	Emanuele Crisostomi,
	Giovanni Russo,
	and~Robert~Shorten
	\thanks{Y. Gu and M. Liu and R. Shorten are with the School of Electrical, Electronic and Communications Engineering, University College Dublin, Ireland (e-mail: yingqi.gu@ucdconnect.ie; mingming.liu@ucd.ie; robert.shorten@ucd.ie).}
		
	\thanks{J. N. Sawaya is with the Ivey Business School, University of Western Ontario, 1255 Western Rd, London, Canada (e-mail: jnaoum-sawaya@ivey.ca).}	
		
	\thanks{E. Crisostomi is with the Department of Energy, Systems, Territory and Constructions Engineering, University of Pisa, Italy (e-mail: emanuele.crisostomi@unipi.it).}
	
	\thanks{G. Russo is with IBM Research, Dublin, Ireland (e-mail:grusso@ie.ibm.com).}}

\markboth{Journal of \LaTeX\ Class Files,~Vol.~6, No.~1, January~2007}%
{Shell \MakeLowercase{\textit{et al.}}: Bare Demo of IEEEtran.cls for Journals}

\maketitle

\begin{abstract}
Electric Vehicles (EVs) and Plug-in Hybrid Electric Vehicles (PHEVs) are increasingly being seen as a means of mitigating the pressing concerns of traffic-related pollution. While hybrid vehicles are usually designed with the objective of minimising fuel consumption, in this paper we propose engine management strategies that also take into account environmental effects of the vehicles to pedestrians outside of the vehicles. Specifically, we present optimisation based engine energy management strategies for PHEVs, that attempt to minimise the environmental impact of pedestrians along the route of the vehicle, while taking account of route dependent uncertainties. We implement the proposed approach in a real PHEV, and evaluate the performance in a hardware-in-the-loop platform. A variety of simulation results are given to illustrate the efficacy of our proposed approach. 
\end{abstract}


\IEEEpeerreviewmaketitle

\section{Introduction}

With the increasing concerns over air quality issues arising from traditional internal combustion engine (ICE) vehicles, plug-in vehicles are seen worldwide as an efficient and effective means of transportation that may help in addressing pollution related challenges. Plug-in Hybrid Electric Vehicles (PHEVs), in particular, are becoming increasingly popular with the general public. Traditionally, such vehicles have two modes of operation: a fully electric mode, and a hybrid mode, the latter of which is designed by the manufacturer to maximise fuel efficiency \cite{stockar2011energy,barsali2004control, sciarretta2004optimal, chau2002overview,amjadi2010power}. Recently, several authors \cite{schlote2013cooperative, IJCSPONGE, naoum2016smart} have suggested exploring the actuation possibilities in such vehicles, namely to automate the on/off switching of the fully electric mode, to address not only fuel efficiency but also pollution issues in urban areas. Specifically, in \cite{schlote2013cooperative}, a number of control strategies were implemented on PHEVs to regulate traffic-related pollution in an urban environment. This was achieved by considering PHEVs as power-split devices and electric energy was used to keep the pollutant level always below a safety pollutant threshold. In particular, the authors had formulated these problems as utility maximisation problems, in terms of different notions of fairness, and had addressed them in a distributed resource-allocation framework using ideas from Internet congestion control. These ideas are further explored for a network of vehicles in \cite{mahsaITS,hausler2014framework} by formulating a constrained optimisation problem. In a similar fashion, for a single vehicle in \cite{guardiola2016adaptive} the authors introduced an online strategy to adapt the engine calibration to the driving conditions continuously, with the final objective of minimising fuel consumption while fulfilling some emission limits for ICE vehicles. A key contribution of this latter work is accounted for the complexity of the ICE in the optimisation problem. Our objective in this present work is to further investigate these ideas. Specifically, we wish to make use of the fact that there are many data feeds that are available in real time in our cities that give reliable information concerning the density of pedestrians. For example, twitter feeds, mobile phone mobility data, and other historical data, all reveal this information. Given this information, and a budget on available electrical power, we wish to orchestrate switching between fully electric mode, and hybrid mode, so that the impact on pedestrians is kept to a minimum. We argue that such data should be, and perhaps must be, incorporated into engine management strategies, so that the on/off switching of the fully electric mode can be orchestrated to minimise the environmental impact of the vehicle on the population. \newline

The current paper extends our previous work in a number of ways. In particular, the work in the present paper builds on \cite{schlote2013cooperative, IJCSPONGE, naoum2016smart, SAMITS}. In \cite{schlote2013cooperative}, we had proposed using feedback control theory to regulate pollution level in a geo-fenced urban area. This was achieved by orchestrating the switching into fully electric mode of a network of vehicles using distributed stochastic algorithms. In \cite{IJCSPONGE,naoum2016smart}, distributed engine management strategies for a fleet of hybrid vehicles were proposed to optimally manage a budget of energy. A severe limitation of this work was that the routes travelled by each vehicle were assumed to be known a priori. This latter issue is addressed in a different context in \cite{SAMITS} in which Markov-decision-based route-prediction engines are proposed and validated \footnote{See \url{https://goo.gl/fpqdt3.} for applications of this engine.}. Our present work uses elements of all four papers to create a framework in which a meaningful optimisation can be formulated with a view to protect pedestrians. Thus, the main contribution of this paper is to propose a new paradigm that manages the way in which a PHEV discharges its limited battery with response to the population density across various routes that may be travelled by a vehicle during a particular journey. This work goes beyond our previous work and other work in the literature by incorporating the following features into the engine management strategies.  \newline

\begin{itemize}
	
	\item [A.] A route prediction engine is designed which, given historical data of the driver, estimates the probability of the routes likely to be travelled by the driver.

	\item [B.] Given A, and an energy budget, online engine management optimisation strategies are proposed, that are based on the distribution of the population over all expected travel routes, as characterised by the route prediction engine.

	\item [C.] Our proposed engine management strategies have been implemented in a real PHEV, and validated using our hardware-in-the-loop (HIL) platform. \\
	
\end{itemize}

The remainder of this paper is organised as follows. Related works are reviewed in Section \ref{relatedwork}. Notation and the problem statement are given in Section \ref{problemformulation}. The system architecture and the optimisation are discussed in Section \ref{system}, where we calculate the probability of each road segment of routes to formulate our optimisation problems. In particular, we note here that a similar approach, by using Markov chain model, can also be applied to calculate the probability as required. For completeness, we also present a high level description on the Markov model in our context in the Appendix section. Implementations of the proposed strategies in the Simulation of Urban MObility (SUMO) package and the HIL platform are presented in Section \ref{simulation}. Simulation results from SUMO and the HIL platform are discussed in Section \ref{discussion}. The limitations of our method and future extensions are remarked in Section \ref{limitations}. Finally, a brief conclusion is presented in Section \ref{conclusion}.

\section{Related Work} \label{relatedwork}

Following the work \cite{schlote2013cooperative}, related work was then developed in \cite{IJCSPONGE,naoum2016smart}. The key idea of such papers was to regulate the energy consumption of EVs on road traffic in a coordinated manner,  using forecasts of available future energy, in order to balance demand and supply and facilitate the adoption of demand side management strategies. Specifically,  by considering the knowledge of both vehicles and energy available at the next charging period, vehicles could control energy consumption along their routes, in a manner that some desired performances indexes of all vehicles could be optimised. These initial ideas are also explored for a network of vehicles in \cite{mahsaITS,hausler2014framework,knorn2011result}. In addition, the authors of \cite{guardiola2016adaptive} have investigated the optimal control strategies to continuously adapt an engine calibration of an ICE vehicle for minimising the fuel consumption while keeping a limited amount of the pollutant emissions over an unknown driving cycle. The optimisation problem was formulated by taking into account the driving behaviours of a given driver and the varying pollutant limits depending on the location of the vehicle and other boundary conditions. The uncertainty of the driving cycle was estimated by using stored historical engine speed and torque demands of the vehicle in a probability manner. Our work is also related to conventional management strategies for hybrid electric vehicles. There are a number of interesting surveys and papers on this mature topic, see \cite{stockar2011energy,chau2002overview,amjadi2010power,barsali2004control,sciarretta2004optimal}. Most of this work focuses on minimising fuel consumption.

\section{Problem Statement}
\label{problemformulation}

Our objective in this paper is to develop an engine management strategy that takes into account the density of pedestrians along a particular route. To do this, we use a relatively simple strategy that does not take into account atmospheric dispersion models for pollutants, nor does it take into account weather or topology information from urban environments. Rather, we take the view that when a vehicle is in polluting mode (hybrid), the probability that it is causing harmful damage is proportional to the population density in that given location. Thus, our strategy is to manage the switching of the engine in order to minimise the impact of pollution on pedestrians; namely, to recommend vehicles to drive in EV mode in high density areas.\newline 

In what follows we will make use of the probability on each segment of routes to formulate the uncertainty that is central to our optimisation formulation; namely, driver intention. 
For this purpose, we define a route $\mathcal{R}$, or a journey, as a sequence of road segments $r_i$ from an origin to a destination point, i.e., $\mathcal{R}=\left\{r_1, r_2, ..., r_N\right\}$, where $N$ is the number of road segments of a specific route. We then define a road segment as the part of a road that connects two consecutive junctions or, if this is too long (e.g., with a density of people that greatly varies along such a part of the road), we may assume that one same road segment could not exceed some fixed length (e.g., 500 meters). Other definitions of road segments may be however used as well, if more convenient (e.g., fixed length). Note that according to the previous definition, a road segment could belong to more than one route. Also, the same origin and destination point may be connected through different routes, if different sequences of segments can be taken to get to the (same) destination. \\ 

\section{System Architecture}
\label{system}

\begin{figure}[htbp]
	\begin{center}
		{\includegraphics[width=3.3in, height = 1.8in]{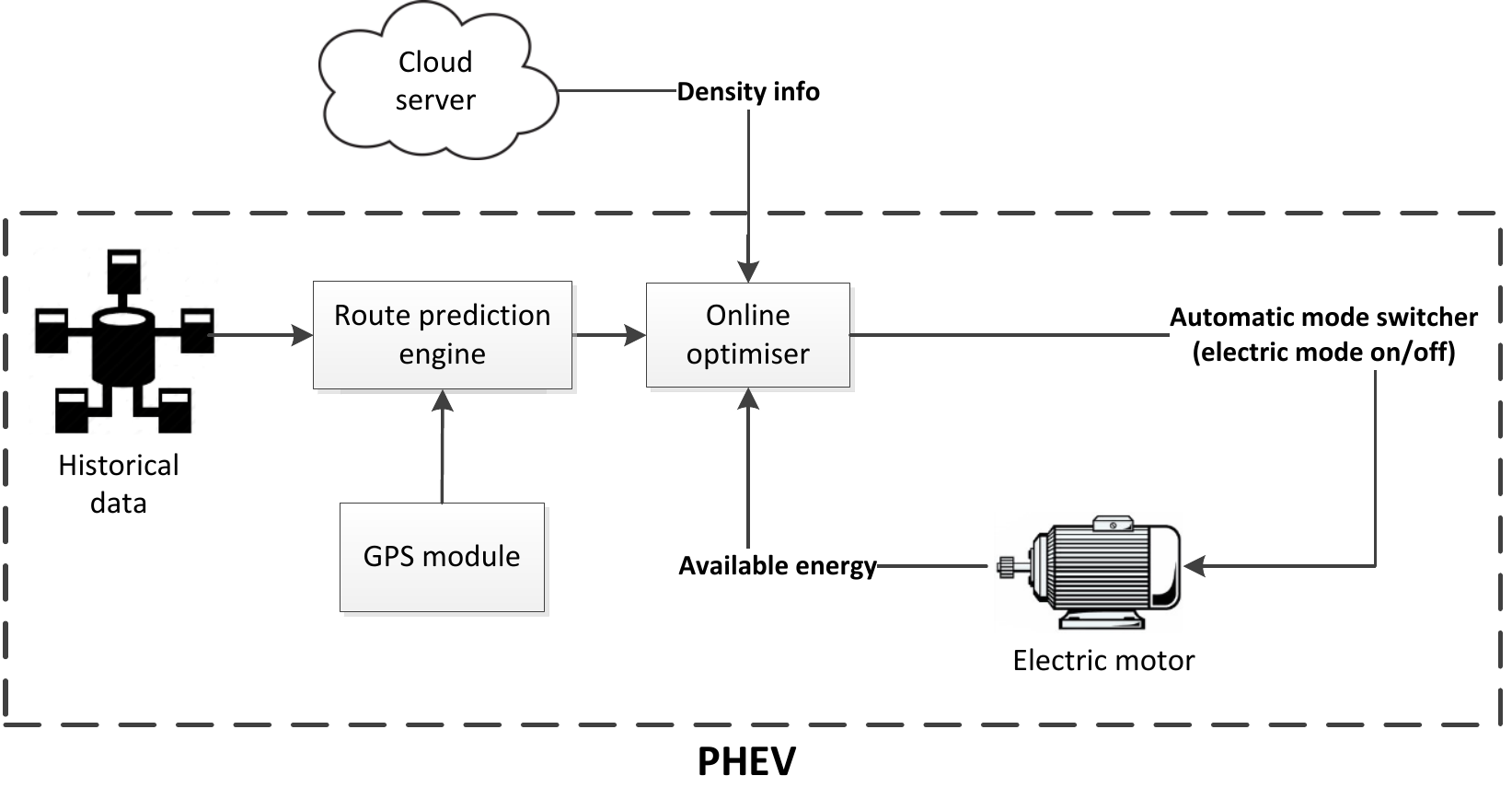}}
		\caption{A schematic diagram of the system architecture.}
		\label{systemModel}
	\end{center}
\end{figure}

Our proposed system architecture is shown in Fig. \ref{systemModel}. The system requires inputs from four functional blocks: 1. historical data of the vehicle (history of routes travelled, energy consumption along these routes)； 2. a route prediction engine that tries to anticipate the driver intention; 3. an online optimiser that controls the engine management; 4. and a cloud server that informs the vehicle of likely pedestrians density. Based on this architecture, the pedestrian-aware engine management strategies operate as follows.

\begin{itemize}
	
	\item[A.]  We assume that a PHEV has an available energy budget for a particular journey. In practice, the energy budget can be estimated from historical energy consumption patterns and the vehicle's current state of charge (SOC). 
	
	\item[B.]  We use the route predictor to obtain the probability distribution over all routes likely to be travelled by the driver based on the historical travelling patterns of the driver and the real-time GPS location of the vehicle (for online optimisation).  
	
	\item[C.]  We require that there exists a central agent (e.g., a cloud server) that has access to real-time population density info in a given area where the vehicles may travel.   
	
	\item[D.]  We use this real-time density info as an external input for the PHEV that will subsequently optimise the sequence of engine switching. 
	
	\item[E.]  The switching to different modes is operated by using a dedicated-device on board that is controlled by the vehicle without input from the driver.
	
	\item[F.]  The mode switching of the PHEV is determined after every road segment in an online optimisation manner. In particular, at the end of a road segment the route predictor may want to update its predictions and rule out some possible routes; the remaining available energy budget of the vehicle may be recalculated. Also, a new estimate of the time-varying number of pedestrians in the different road segments may become available.
	
\end{itemize} 


In the following we outline the details of such functional blocks.  



\subsection{Route prediction engine}
The objective of the route prediction engine is to predict the unknown route of the driver, and, more specifically, also to predict the probability of taking every possible road segment. In particular, consider a generic road segment $r_i$ where the vehicle is currently along with, and assume that this segment belongs to a number of routes, say $\mathcal{R}_1,\ldots, \mathcal{R}_k$. Moreover, we denote by $N_t$ the total number of times that route $\mathcal{R}_t$ was taken in the past. Then, a probability, say $p(\mathcal{R}_t)$ is computed for each of these routes as follows:

\begin{equation}
p(\mathcal{R}_t) = \frac{N_t}{\sum_{j=1}^k N_j}, \forall t \in \left\lbrace 1,2,\ldots, k \right\rbrace.
\end{equation}

In particular, if a route does not pass by $r_i$, then the probability that is on the actual route is considered zero. Also, if only one route passes by $r_i$ then its probability is considered 1. Then, the probability $p_s^i$ of taking segment $r_s$ in the future is simply computed as the sum of the probabilities of all routes that contain that road segment. In the notation $p_s^i$, the superscript $i$ reminds that the probability of a route, and consequently, of a road segment within that route, depends on the specific road segment $r_i$ where the vehicle is travelling at the time when a new prediction of the route is performed.\newline

{\noindent \bf Remark: } In our model, we implicitly assume that a car has the ability to store routes travelled in the past, and the number of times each route was taken. For instance, we could realistically assume that each vehicle can store up to 100 routes. After this number, any new route replaces an older rare route (e.g., one route that was taken only once). Thus, storing route information in the manner described here can be challenging in terms of data storage. A compact method of storing driver route choices is to embed driver's intention in a Markov chain. For simplicity, we do not describe the Markov chain approach in detail here, but some details and useful references are given in the Appendix. \\

\subsection{Optimisation}
In this section we formulate our online energy management optimisation problems. For this purpose, we add the following notation: \textit{(i)} we denote by $\mathcal{S}$ the set of all road segments appearing in all the past routes; \textit{(ii)} we denote by $\overline{e}_{s}$ the expected energy consumption along the $s$'th segment of $\mathcal{S}$, when travelling in pure electric mode (again, this can estimated from the historical data); \textit{(iii)} we denote by $d_s^i$ the expected number of people along the $s'$th road segment of $\mathcal{S}$ (as mentioned this can be computed using twitter feeds, mobile phone data, or can be simply estimated again from some historical data). The superscript $i$ refers to the fact that we are using the information that is available when the vehicle is along road segment $r_i$; \textit{(iv)} we denote by $E^i_{\textrm{av}}$ an available energy budget when the vehicle is located at the $i'$th segment. For instance, this could simply be the energy left in the battery when the vehicle is along road segment $r_i$; finally \textit{(v)} we denote by $x_s^i \in \left[0, \ 1\right]$ the set of decision variables that we wish to optimally compute. In particular, $x_s^i$ is the percentage of time that the vehicle is in full electric mode on each segment $r_s \in \mathcal{S}$. The superscript $i$ here reminds again that such an optimal prediction is performed when the vehicle is driving on road segment $r_i$, and a new prediction will be performed when the vehicle enters a new road segment, when a new (possibly more accurate) prediction of the route will be performed.\newline

\noindent \textbf{\textrm{Problem 1:}} Mathematically, the first optimisation problem that we are interested in is:
\begin{equation} 
	\label{optproblem}
	\left\{
	\begin {array}{l}
	\underset{x^i_s}{\max} \quad \sum_{s \in S} p^i_s d^i_s \overline{e}_{s} x^{i}_{s}     
	\\
	{\text{s.t.}} ~ \sum_{s \in S} p^i_s \overline{e}_{s} x^{i}_{s}  \leq E^i_{av}\\
	\\ 
	\quad 0 \leq x^{i}_{s} \leq 1
\end{array}
\right.,
\end{equation}
where $S=\left\{i \in \mathbb{N} | r_i \in \mathcal{S}\right\}$ (i.e., the constraint involves all the road segments). Roughly speaking, Problem 1 aims at making the vehicle travel in electric mode more likely when there are more pedestrians. In particular, this is done in a probabilistic manner, due to the uncertainty of the route, by giving more importance to the most likely routes. \newline


\noindent \textbf{Remark:} Problem 1 is solved iteratively every time the vehicle enters a new road segment; in fact, when the vehicle enters a new road segment, it is possible to update the prediction of the route, to possibly update the remaining energy in the battery and the number of pedestrians as well. \newline

\noindent \textbf{Problem 2:}
While Problem 1 refers to an average optimisation (averaged over all possible routes with their corresponding probabilities), it may be also of interest to optimise the variables $x^{i}_{s}$ with respect to the most energy-consuming route. In this case, the optimisation problem is solved with respect to the worst-case scenario (i.e., energy will be instantaneously allocated taking into account that the actual route might be the most energy demanding one). In this case, the mathematical problem may be formulated as:
\begin{equation}
\label{newconstraint}
\sum_{s \in R} \overline{e}_{s}x^{i}_{s}  \leq E^i_{av} , ~ \text{for all routes}, \\
\end{equation}
where $R=\left\{i \in \mathbb{N} | r_i \in \mathcal{R}\right\}$ (i.e., the constraint involves all the road segments belonging to a given route, and must then hold for all routes). This formulation guarantees that the available battery level is never exceeded, irrespective of the driver choices. As we shall see later, such a worst-case scenario usually gives rise to a more conservative energy consumption pattern. \\



\noindent \textbf{Example:} We now provide a simple example to clarify our approach. In particular, we assume that one vehicle starts from the beginning of road segment $r_1$ that in the historical database appears in three different routes, as below:
\begin{itemize}
	\item Route 1: $r_1 \rightarrow r_2 \rightarrow r_3$, this route was taken 100 times;
	\item Route 2: $r_1 \rightarrow r_2 \rightarrow r_4$, this route was taken 200 times;
	\item Route 3: $r_1 \rightarrow r_5 \rightarrow r_4$, this route was taken 400 times.
\end{itemize}
Then $\mathcal{S}=\left\{r_1, r_2, r_3, r_4, r_5\right\}$. Segments $r_3$ and $r_5$ appear only in route $1$ and $3$ respectively, so their probabilities $p^1_3$ and $p^1_5$ correspond to the probabilities of the corresponding routes (i.e., $1/7$ and $4/7$ respectively). On the other hand, probability $p^1_2$ is $3/7$ (as both routes 1 and 2 lead to road segment $r_2$) and the probability $p^1_4$ is $6/7$ (as both routes 2 and 3 lead to road segment $r_4$). In this case, we have that the objective function of the optimisation problem (first equation of \eqref{optproblem}) is
\begin{equation}
\begin{split}
d^{1}_{1} \overline{e}_{1} x^{1}_{1} + \frac{3}{7} d^{1}_{2} \overline{e}_{2} x^{1}_{2} + \frac{1}{7} d^{1}_{3} \overline{e}_{3} x^{1}_{3} + \frac{6}{7} d^{1}_{4} \overline{e}_{4} x^{1}_{4} + \frac{4}{7} d^{1}_{5} \overline{e}_{5} x^{1}_{5},
\end{split}
\end{equation}
where the objective is to compute the optimal values of $x$'s, given the knowledge of the number of pedestrians and the expected energy consumption along each road segment.
Also, the energy constraint is then
\begin{equation}
\overline{e}_{1} x^{1}_{1} +  \frac{3}{7} \overline{e}_{2} x^{1}_{2} + \frac{1}{7} \overline{e}_{3} x^{1}_{3} + \frac{6}{7} \overline{e}_{4} x^{1}_{4} + \frac{4}{7} \overline{e}_{5} x^{1}_{5} \leq E^{1}_{av},
\end{equation}
if we wish to solve Problem 1, and
\begin{equation}
\left\{\begin{array}{l}
\overline{e}_{1} x^{1}_{1} +  \overline{e}_{2} x^{1}_{2} + \overline{e}_{3} x^{1}_{3} \leq E^{1}_{av} \\
\\
\overline{e}_{1} x^{1}_{1} +  \overline{e}_{2} x^{1}_{2} + \overline{e}_{4} x^{1}_{4} \leq E^{1}_{av} \\
\\
\overline{e}_{1} x^{1}_{1} +  \overline{e}_{5} x^{1}_{5} + \overline{e}_{4} x^{1}_{4} \leq E^{1}_{av} \\
\end{array}\right.,
\end{equation}
if we address the worst-case scenario of Problem 2. \newline

\noindent \textbf{Remark (Networked Vehicles):} Here we briefly remark that, our current framework can be easily extended in a variety of ways to incorporate the behaviours of networked vehicles for minimising the group pollutants to pedestrians. For instance, given a fleet of PHEVs having similar travelling route maps in a fixed city area, one may wish to adopt our proposed framework for all vehicles optimally switching their engine modes while considering the impact on traffic flow on all possible routes of all vehicles. In practice, this problem can be easily formulated for each PHEV in the form of \eqref{optproblem} which gives rise to our Problem 3. \\

\noindent \textbf{Problem 3:} Formally speaking, the problem of the networked vehicles can be formulated in a distributed manner, namely each PHEV solves its own problem individually. Mathematically, we are interested in solving:   
 
\begin{equation} 
\label{optproblemNetwork}
\left\{
\begin {array}{l}
\underset{x_s^i}{\max} \quad \sum_{s \in S}  p_{s}^{i}d^{i}_{s}\overline{e}_{s} x^{i}_{s} f_{s}^{i}    
\\
{\text{s.t.}} ~ \sum_{s  \in S} p_{s}^{i}\overline{e}_{s}x^{i}_{s}  \leq E^i_{av}\\
\\ 
\quad 0 \leq x^{i}_{s} \leq 1.
\end{array}
\right.
\end{equation}
where $f_s^i$ denotes the traffic flow on road segment $s$ relative to all other road segments when the vehicle starts from the beginning of the $i$'th segment.

\section{SUMO and HIL simulations}\label{simulation}

In this section, we mainly introduce our simulation set-up in SUMO and present our algorithm implementations in the hardware-in-the-loop (HIL) platform embedded with a real car. First note that a complete description of the HIL platform is presented in \cite{griggs2015large}. In this work, we shall test our applications based on this platform with proper modifications and extensions for specific experimental purposes. Here we repeat some contents from  \cite{griggs2015large} for readers' conveniences.

\subsection{Simulation set-up in SUMO} \label{SUMOsetup}

\begin{figure}
		\includegraphics[width=.5\textwidth,height = 2.8in, right]{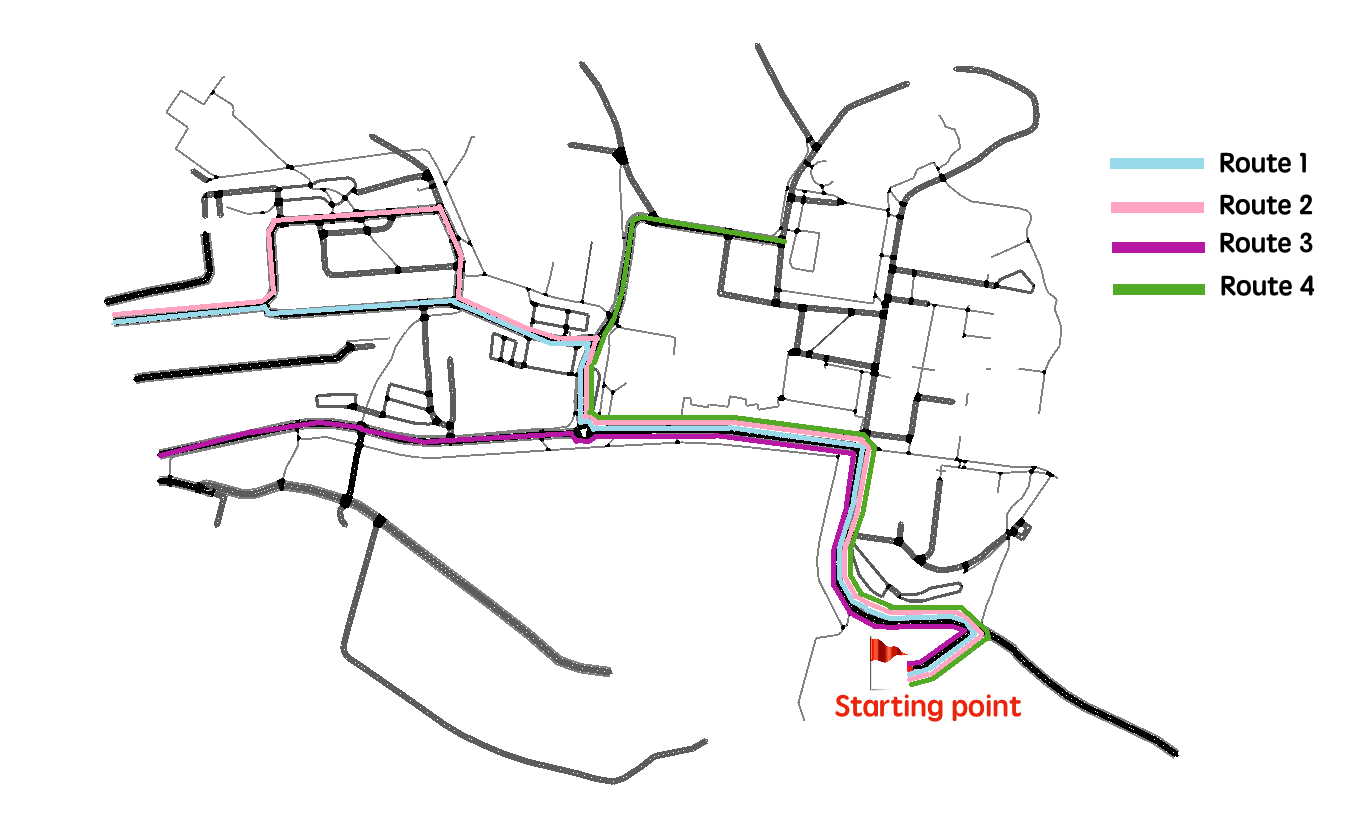}
		\caption{Four real routes used in the field experiments of our test PHEV.}
		\label{UCDmap}
\end{figure}

\begin{figure}[htbp]
		\includegraphics[width=.5\textwidth, height = 1.8in, right]{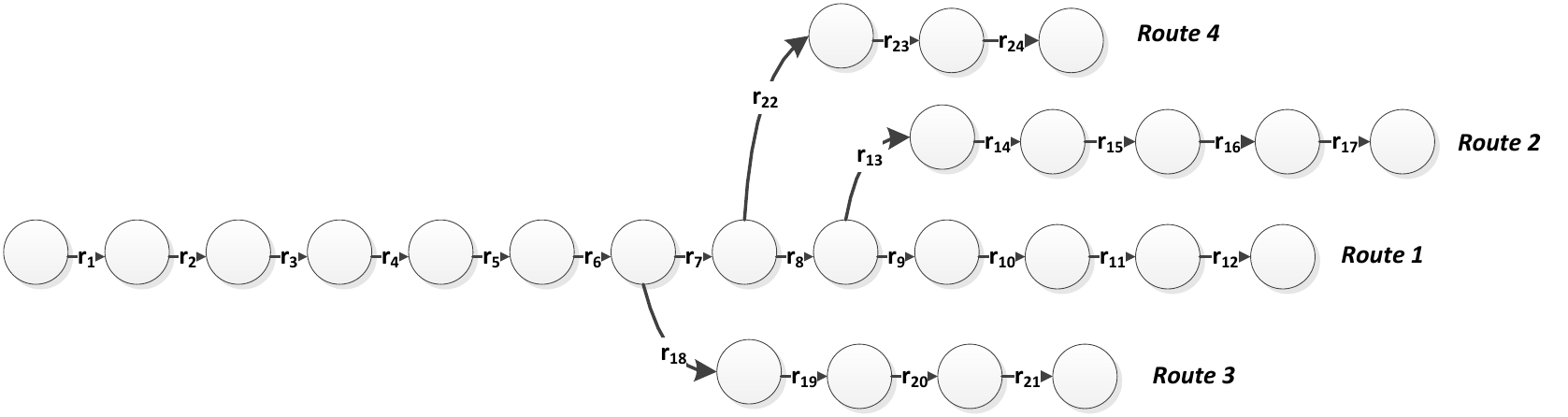}
		\caption{A probability model including four real test routes in UCD campus. Note: 1. Each circle represents the beginning of one (or more) road segment(s), where the vehicle solves the online optimisation problem; 2. Length of the arrow is for indicating purpose only and it does not represent the real length of a road segment.}
		\label{markovexampleUCD}
\end{figure}

In this section, we evaluate the performance of our algorithm in a realistic traffic scenario in the University College Dublin (UCD) campus, where the mobility of our test PHEV (Prius) is simulated in the popular traffic simulator software package SUMO \cite{SUMO}. The road network of the campus is imported from OpenStreetMap \cite{openstreetmap} and loaded in SUMO for further simulation and analysis. Specifically, we assumed that our test PHEV (Prius) had four route records in the UCD campus from its vehicle database. The road network and the corresponding route info is shown in Fig. \ref{UCDmap}. Note that we also assumed that all routes of the vehicle shared the same starting point but ended up with different exits (or junctions), which represent different destinations that the vehicle might drive out of the campus. It is worth noting that all routes have some road overlaps with each other; and this can be seen clearly from its probability model pictured in Fig. \ref{markovexampleUCD}.

\subsection{Real car implementation}

\begin{figure}[htbp]
	\begin{center}
		{\includegraphics[width=3.5in, height = 2.4in]{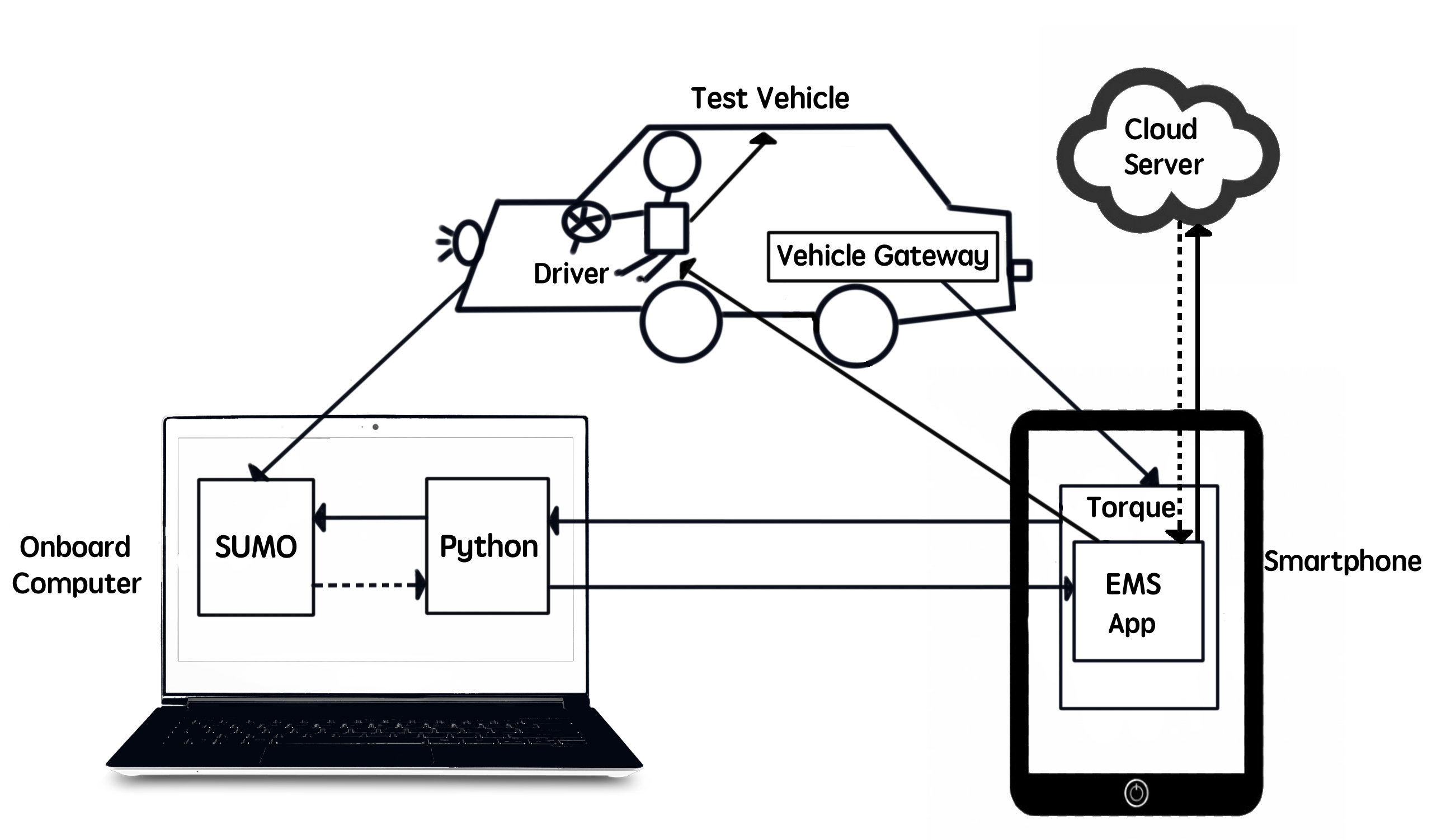}}
		\caption{A schematic digram of the real car implementation. Note that in this figure EMS is short for Energy Management System, and two dash lines represent two different ways that Python can obtain real time density info of pedestrians for online optimisation. In this work we simply adopt the interface from SUMO as an external input.}
		\label{schematic}
	\end{center}
\end{figure}

In this section, we introduce our real car implementation using the HIL platform. A schematic diagram of our real car implementation is shown in Fig. \ref{schematic}. The system mainly consists of a test vehicle, an onboard computer, a smartphone, and a cloud server. 
\newline

\begin{figure}[htbp]
		\begin{center}
			{\includegraphics[width=3.0in, height = 2.3in]{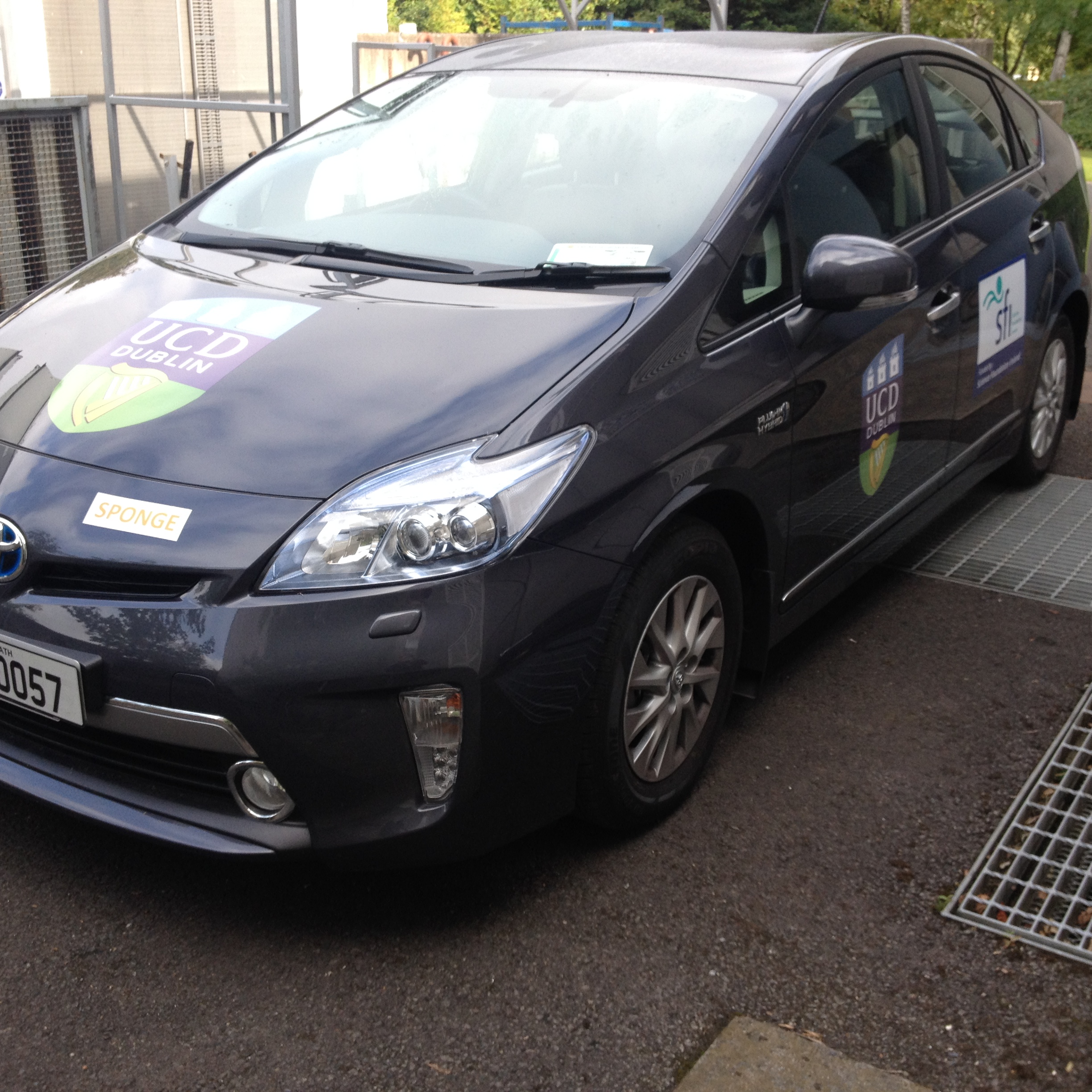}}
			\caption{Field-test vehicle: 2015 Toyota Prius.}
			\label{prius}
		\end{center}
	\end{figure}

     \textit{A. Test Vehicle:} The test vehicle we used for our field experiments is a 2015 Toyota Prius VVTi 1.8 5DR CVT Plug-in Hybrid vehicle and is pictured in Fig. \ref{prius}. One of the advantages of using the Prius as our field-test vehicle is because of its flexible engine management system, which allows us to operate the vehicle in fully electric mode and hybrid mode. It is this degree of freedom that allows us to explore the optimality of switching in response to pedestrians. To facilitate the automation between different mode switch (e.g., from fully electric mode to hybrid mode), we designed a dedicated mechanical interface (like a ``Finger'') which overrides the manual EV button in the vehicle. In our application, this interface is triggered by receiving Bluetooth control signals from a smartphone. The control signal is transmitted from phone to the ``Finger'', after every fixed driving distance of the Prius (e.g., 100 meters), based on the optimal algorithm output calculated from the onboard computer running with SUMO. \newline
	
    Finally, we constructed a special-purpose hardware to permit communication between a smartphone and the controller area network (CAN) bus. The Prius provides a CAN access on the vehicle diagnosis {\em On Board Diagnosis II (OBDII) interface} and the OBDII interface device that we used is the Kiwi Bluetooth OBD-II Adaptor by PLX Devices\footnote{PLX Devices Inc., 440 Oakmead Parkway, Sunnyvale, CA 94085, USA.  Phone:  +1 (408) 7457591.  Website:  \url{http://www.plxdevices.com}}. Our hardware module acts as a gateway between this CAN interface and the smartphone. The module is directly connected to the CAN and to the smartphone via Bluetooth. \newline
        
    \vspace{-0.5cm}    
    \textit{B. Onboard Computer:} In our application, the optimisation algorithm is written in Python and is implemented on an onboard computer running with SUMO. SUMO is a mobility traffic simulator that we used to simulate the mobility of the test vehicle in our scenarios. We used SUMO to load the map imported from OpenStreetMap, and used its handy interface, namely TraCI (short for Traffic Control Interface) \cite{b4}, to retrive online states info of the real vehicle from the smart phone via Python and control the behaviours of the vehicle in SUMO on the fly. Specifically, we used SUMO to model the behaviours of pedestrians walking in the campus as well, and based on this to provide population density info to Python as an external input. It is worth mentioning that a similar function can be achieved by using a cloud server, and we shall give brief comments for doing this in the Cloud Server section. \newline

    Based on all data collected, which includes real time data, such as Prius's GPS location, battery level, speed, distance and pedestrians data, as well as historical travelling data from the driver, Python implements the algorithm and determines the optimal duration of the vehicle that the electric engine mode should be engaged for pedestrian-aware driving. \newline

\begin{figure}[htbp]
    	\begin{center}
    		{\includegraphics[width=3.4in, height = 2.4in]{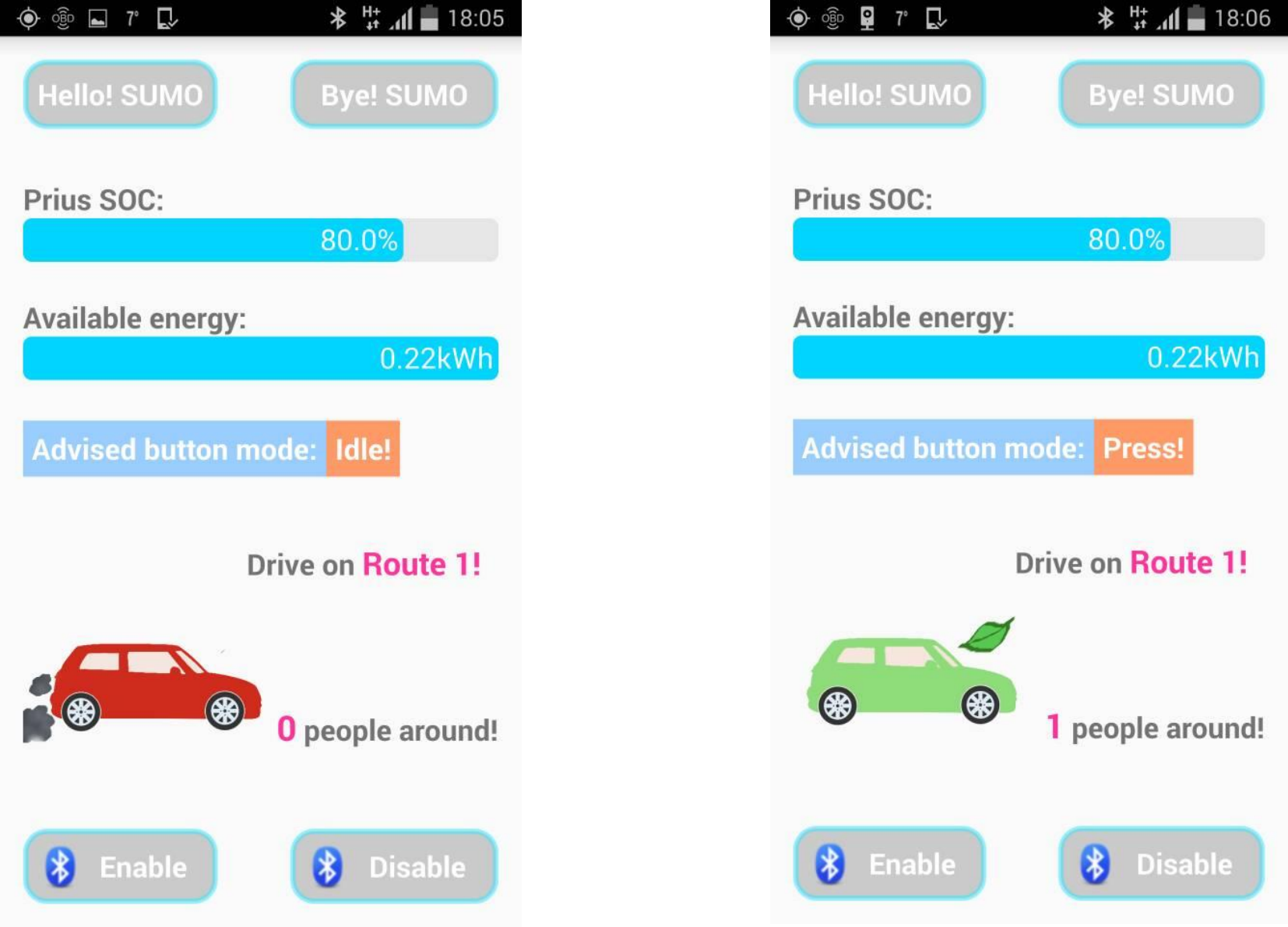}}
    		\caption{App GUI design for the field experiments. Figure on the left suggests the PHEV should be operating in hybrid mode and the figure on the right advises the PHEV should be switched to electric mode.}
    		\label{AppGUI}
    	\end{center}
    \end{figure}

    \textit{C. Smartphone:}  In our architecture, the smartphone is required to collect data transmitted from the vehicle gateway (and from the cloud server if it is required) and forward such data to the onboard computer for online optimisation.  From this viewpoint, the smartphone is only acting as a relay in our implementation system. In this work we developed a specific app on a Samsung Galaxy S IV (model no. GT-I9500) smartphone running the Android KitKat operating system (version 4.4.2) to demonstrate essential vehicle info to the driver. The GUI of the App we designed is shown in Fig. \ref{AppGUI}.  \newline
    
\begin{figure}[htbp]
    	\begin{center}
    		{\includegraphics[width=3.4in, height = 2.4in]{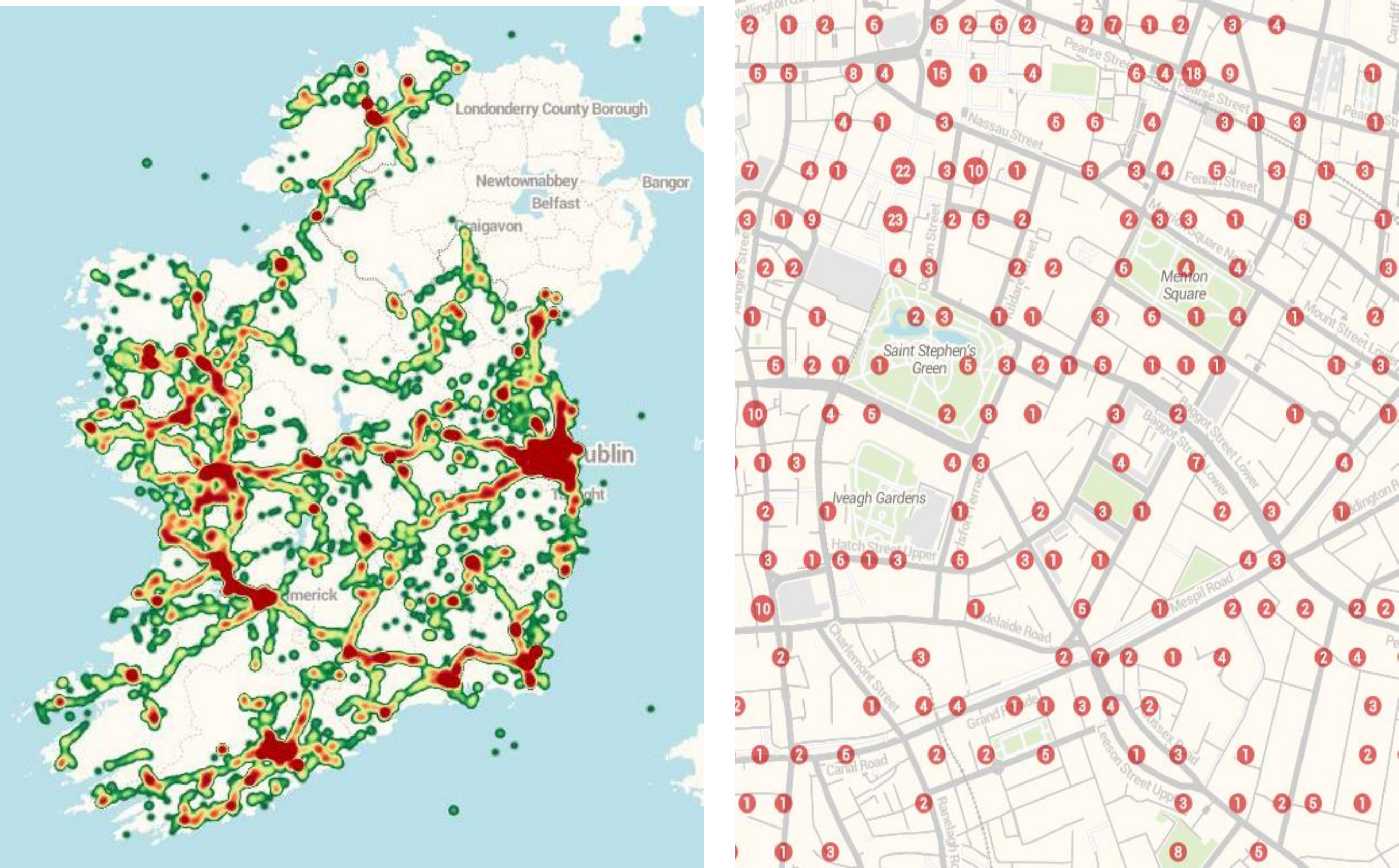}}
    		\caption{Heat map of the density distribution of mobile phone users in Ireland (left) and the cluster map of the density in areas near Dublin city centre (right) on a typical working day, where the red colour in the heat map indicates higher density than the green one and the number in cluster map indicates the number of signal measurements received from the cell towers. Note: data is collected via OpenCellID.org API: http://opencellid.org/api on 5th of July, 2016.}
    		\label{heatmap}
    	\end{center}
    \end{figure}

    \textit{D. Cloud Server:} A cloud server is only required to act as a central agent being capable of sending specific pedestrians info, via realisable communication channels, to smartphone whenever it receives request. In practice, a twitter feed \footnote{\url{http://twitterfeed.com/}} or any applicable third party service supplier (e.g., free mobile station data from OpenCellID \footnote{\url{http://www.opencellid.org}}) can be directly used, or indirectly used, where exact pedestrians data is not available, on behalf of the cloud server to take actions in this step. As a concrete example, a real Ireland-based density distribution for mobile phone users (only those who installed the OpenCellID app) on a specific day is given in Fig. \ref{heatmap}, where the density data across Ireland is filtered from the original data obtained from the OpenCellID database. Finally, we note that it is not required in our design that the cloud server should be capable of collecting all historical travelling data from Prius, and thus this not only reduces heavy computing burdens on the cloud side but also preserves privacy of the driver.

\section{Results and Discussions}\label{discussion}

\begin{figure}[htbp]
	\begin{center}
		{\includegraphics[width=3in, height = 2.4in]{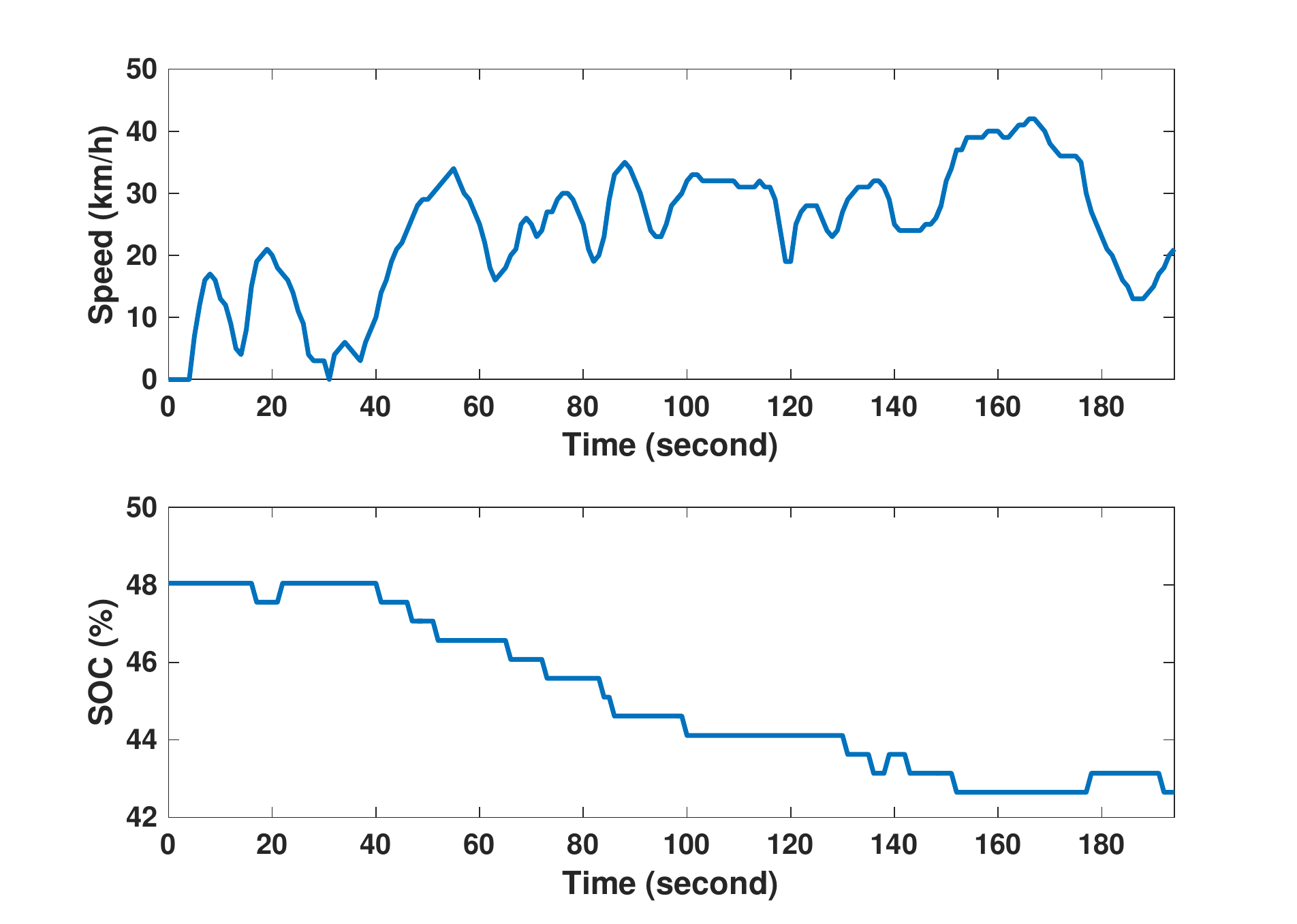}}
		\caption{The speed and SOC of Prius in pure electric mode in one of the field experiments on route 2.}
		\label{speedandsoc}
	\end{center}
\end{figure}

In this section we discuss the simulation results obtained by implementing the proposed engine management strategies on a real car. To begin with, we use the proposed probability model in Fig. \ref{markovexampleUCD} and we define the maximum length of a road segment as 100 meters. To generate multiple historical energy consumption data of the Prius in full electric mode, we randomly distribute the energy on each road segment between 0 and 0.05 kWh according to the estimation from our real experimental data. As an example, Fig. \ref{speedandsoc} shows the real driving speed and the state of charge of the Prius in pure electric mode during one of our field experiments on route 2. Further, we assume a probability of 40\% for the PHEV to travel on route 1, 30\% for route 2, 20\% for route 3, and 10\% for route 4. Finally, we assume that the density of pedestrians along all routes is time-varying and is available on request. \newline

\subsection{Case study: Problem 1}

In this section we illustrate the simulation results obtained by solving optimisation Problem 1. Specifically, we compare different energy management schemes of Prius in three scenarios, while the PHEV was driving on route 1. We assume that the Prius started from the beginning location of segment $r_1$ in all scenarios and has the same initial energy budget 0.22 kWh which corresponds to 5\% of its battery size (4.4 kWh) at the beginning of driving. \newline

In the first scenario, the PHEV solves the optimisation Problem 1 at the beginning of every road segment along route~1. In particular, the $\overline{e}_{s}$ is calculated using real data measured from the Prius averaged over multiple experiments~\footnote{It can also be calculated from a vehicle model and data corresponding to congestion information at different times of the day.}. In the second scenario, instead of averaging, $\overline{e}_{s}$ is taken to be the maximum value observed for a particular road segment over all experimental data. The third scenario, which is included as a benchmark, does not use any optimisation. Instead, the electric mode is engaged in proportion to the maximum observed $\overline{e}_{s}$ along a given route. For clarity, we abbreviate these scenarios as ``Average-Forecast'', ``Max-Forecast'', and ``None-Opt'' in the following. \newline

\begin{figure}[htbp]
	\begin{center}
		{\includegraphics[width=3.2in, height = 2.4in]{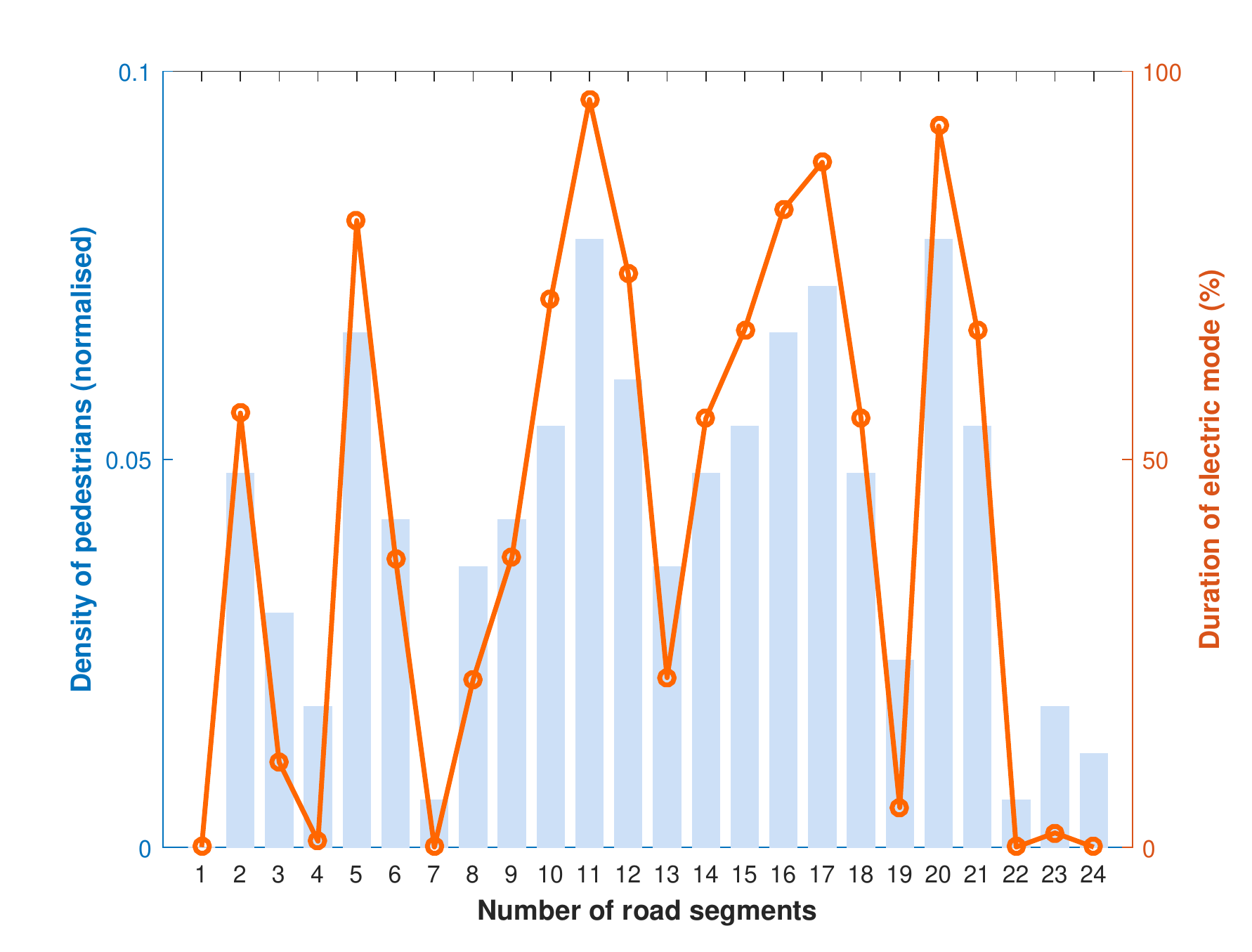}}
		\caption{Comparison of normalised density of pedestrians and the averaged duration of electric mode, with respect to 1000 different initial energy budgets, when the PHEV starts to drive at $r_1$.}
		\label{NewResult1}
	\end{center}
\end{figure}

\begin{figure}[htbp]
	\begin{center}
		{\includegraphics[width=3.2in, height = 2.4in]{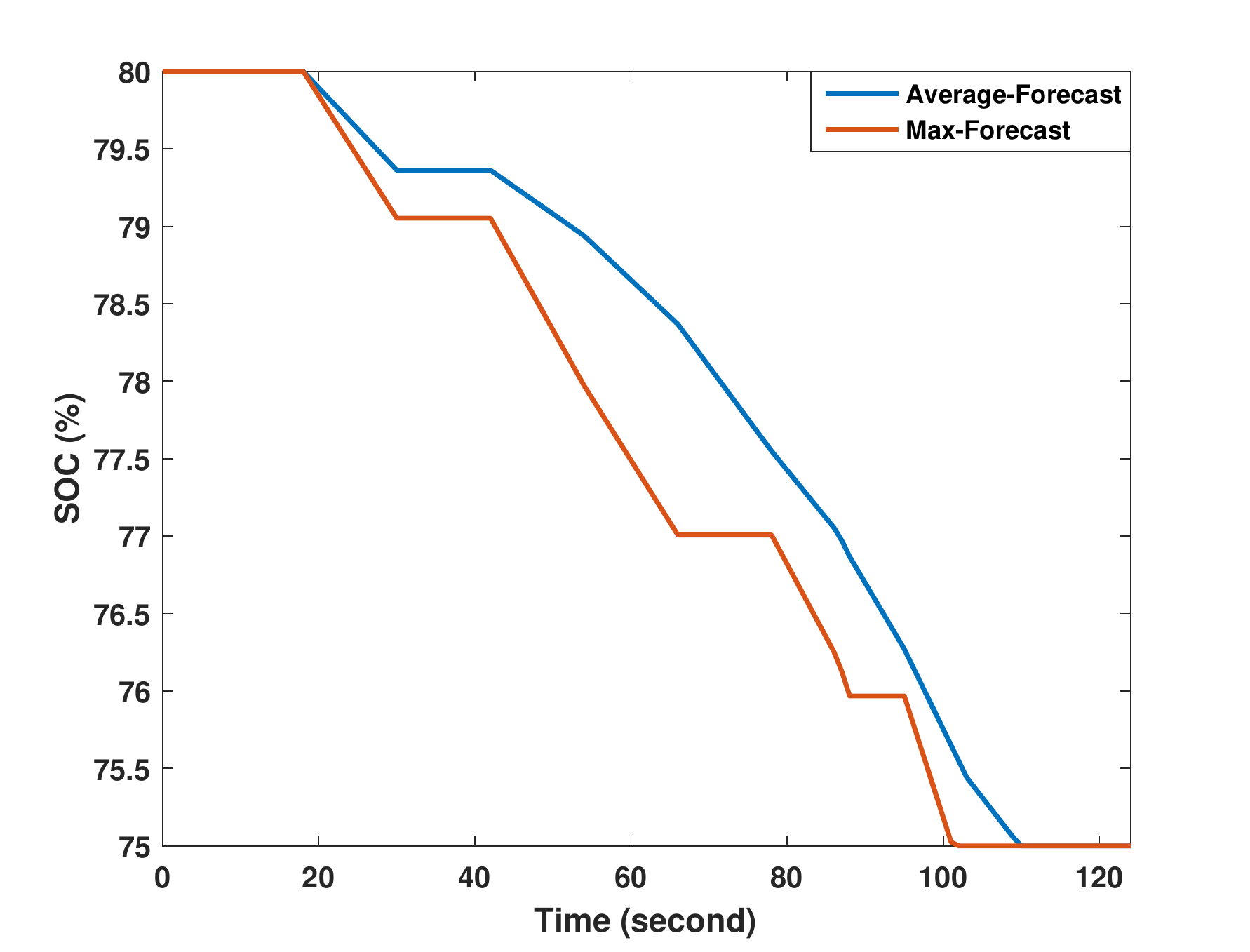}}
		\caption{State of charge of the PHEV in the first two scenarios.}
		\label{NewResult2}
	\end{center}
\end{figure}

\begin{figure}[htbp]
	\begin{center}
		{\includegraphics[width=3.2in, height = 2.4in]{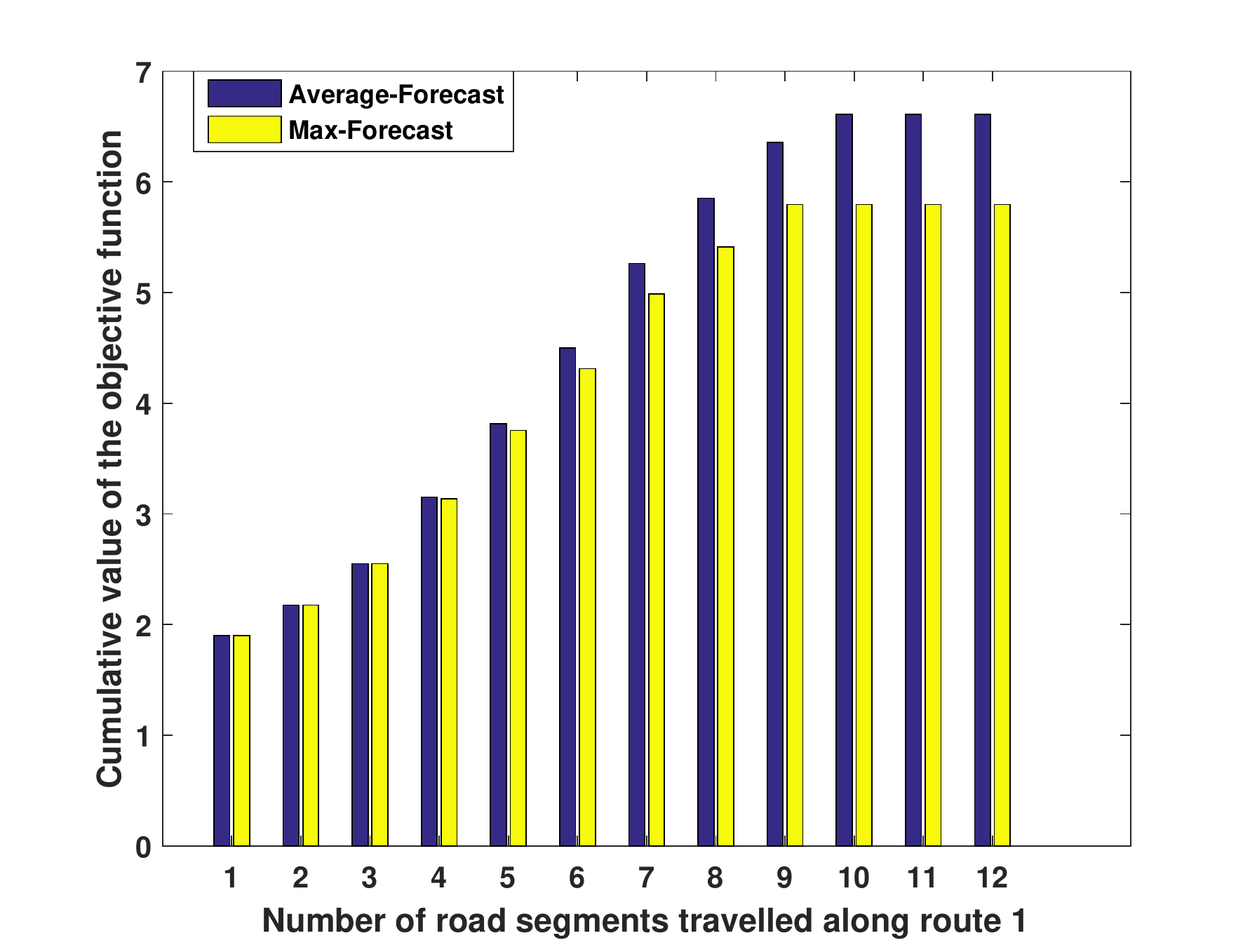}}
		\caption{Comparison of the cumulative value of the objective function in the first two scenarios.}
		\label{NewResult3}
	\end{center}
\end{figure}

\begin{figure}[htbp]
	\begin{center}
		{\includegraphics[width=3.2in, height = 2.4in]{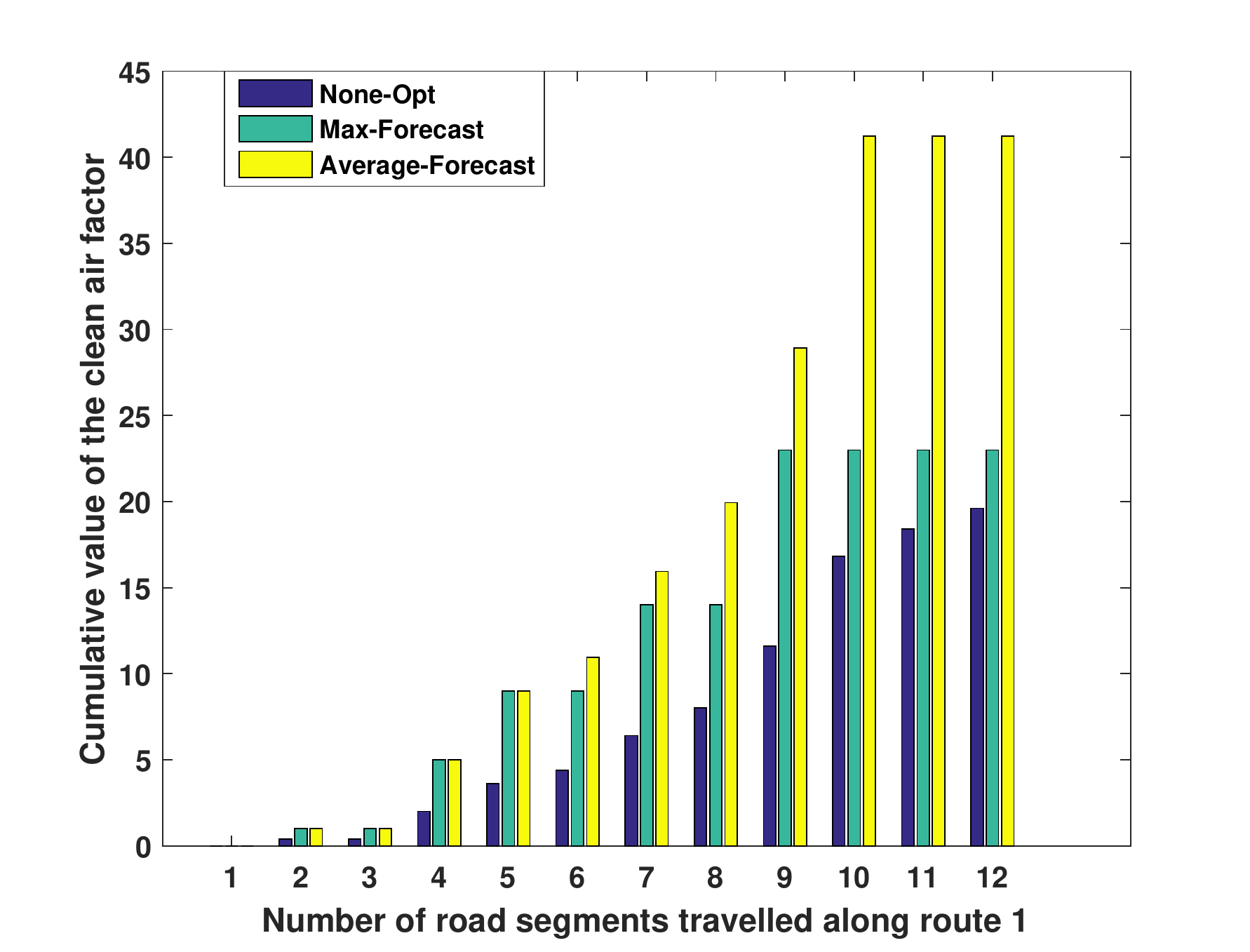}}
		\caption{Comparison of the cumulative value of the clean air factor in all scenarios.}
		\label{NewResult4}
	\end{center}
\end{figure}

Fig. \ref{NewResult1} shows the relation between the normalised population density and the duration of electric mode in the first scenario when the PHEV starts to drive at $r_1$. The duration of electric mode at each road segment is averaged by solving optimisation Problem 1 in 1000 times by linearly spacing the available energy $E^i_{av}$ between 0 kWh and 0.22 kWh (i.e., step size equals $0.22 \cdot 10^{-3}$). The results show that our approach preferentially allocates electric energy to those road segments with higher density of pedestrians.  A comparison of results between ``Average-Forecast'' and ``Max-Forecast'' are shown in Fig. \ref{NewResult2} and Fig. \ref{NewResult3}. Fig. \ref{NewResult2} shows that the state of charge (SOC) of the PHEV decreases more sharply in ``Max-Forecast'' scenario than in the other scenario due to the fact that the largest energy consumption patten of the PHEV was adopted, while it also shows that the proposed method manages the energy consumption of the PHEV in exactly 0.22 kWh (5\% of the battery size) along route 1 in both scenarios. As expected, Fig. \ref{NewResult3} illustrates that the cumulative value of the objective function in ``Average-Forecast'' is indeed larger than ``Max-Forecast''.  Finally, Fig. \ref{NewResult4} compares the value of cumulative clean air factor along route 1 in all scenarios, where the clean air factor is defined as $d_s^{i} x_{s}^{i}$ at each road segment $s$ when the PHEV is in $i$. In there, we observe that a continuous improvement in air quality is achieved by a simple implementation of our proposed strategy, compared to the trivial equal energy allocation scheme (``None-Opt'').

\subsection{Case study: Problem 2}

\begin{figure}[htbp]
	\begin{center}
		{\includegraphics[width=3.2in, height = 2.4in]{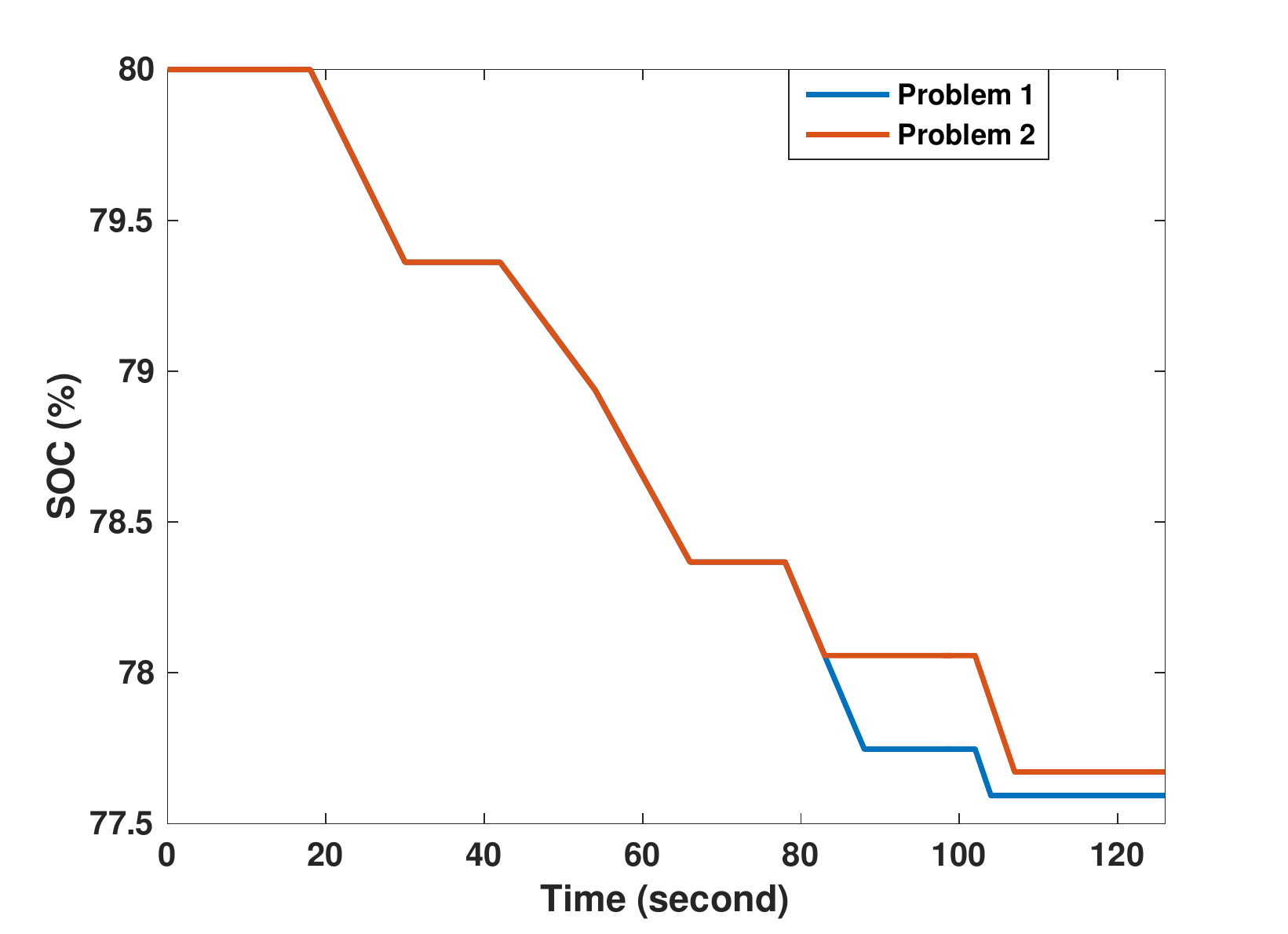}}
		\caption{Comparison of the SOC of the PHEV on Route 3 using expected (Problem 1) and robust (Problem 2) energy constraints.}
		\label{NewResult5}
	\end{center}
\end{figure}

\begin{figure}[htbp]
	\begin{center}
		{\includegraphics[width=3.2in, height = 2.4in]{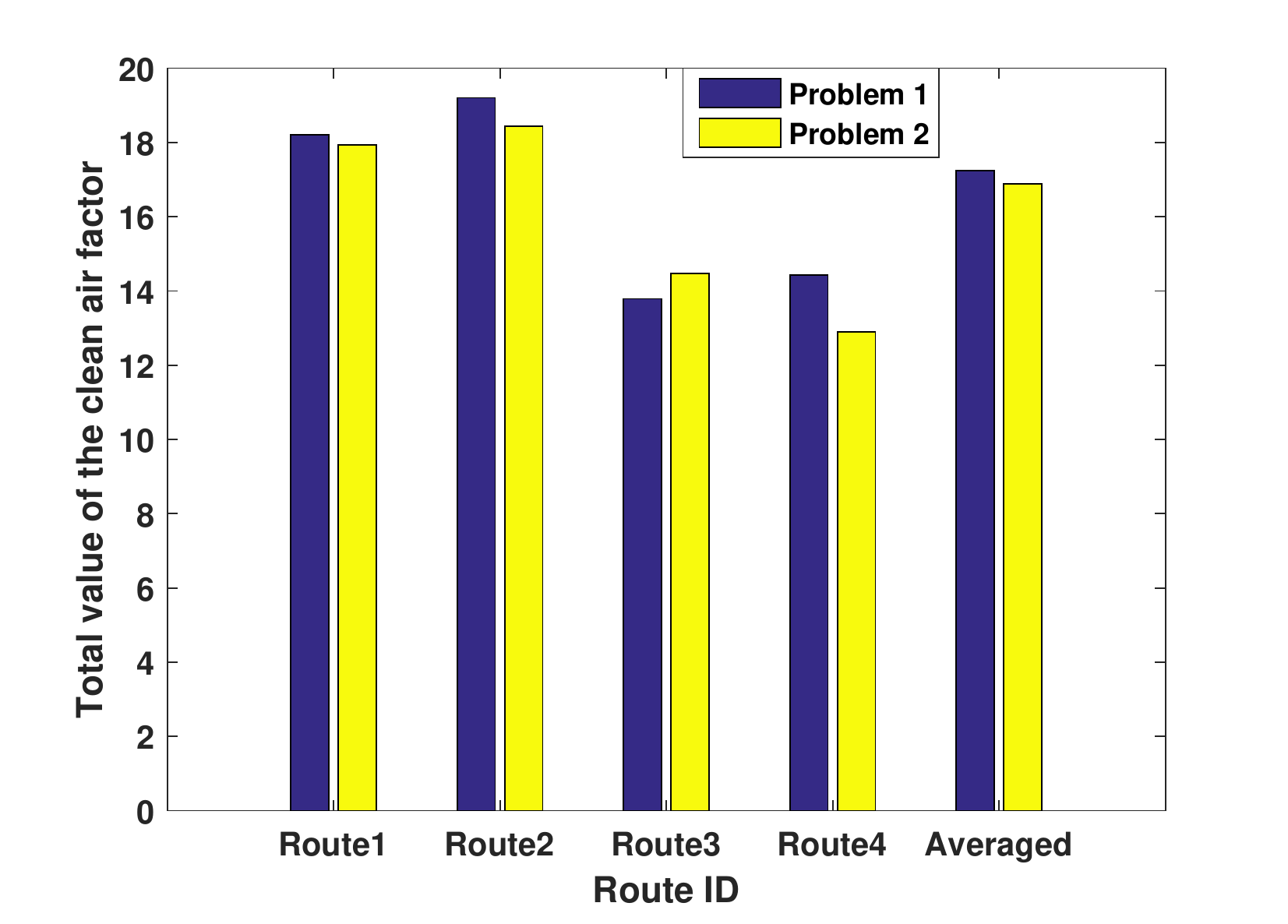}}
		\caption{Comparison of the total clean air factor using expected (Problem 1) and robust (Problem 2) energy constraints on different routes.}
		\label{NewResult6}
	\end{center}
\end{figure}

In this section we illustrate the simulation results obtained by solving optimisation Problem 2. In particular, we compare our results to the first method of solving Problem 1. In this study, we assume that the Prius has 0.1 kWh energy budget (2.5 \% of battery size) before driving and we evaluate the overall clean air factors as if the Prius was driving on 4 different routes starting from the beginning location of $r_1$. Our results are shown in Fig. \ref{NewResult5} and Fig. \ref{NewResult6}. Fig. \ref{NewResult5} shows that the energy consumption of the PHEV is indeed more conservative (after 80 seconds) compared to using Problem 1 on Route 3, which is exactly the most energy-inefficient route in our study. Most importantly, from the last two bars in Fig. \ref{NewResult6} we can see that, on average, the overall air quality evaluated using the expected energy constraint (i.e., the first constraint in \eqref{optproblem}) is better compared to when robust energy constraints are used (i.e., equation \eqref{newconstraint}). However, as shown on route 3 in Fig. \ref{NewResult6}, for the route where the PHEV achieves the worst air quality, the robust energy constraints lead to better air quality.

	\subsection{Case study: Problem 3}
	
\begin{figure}[htbp]
		\begin{center}
			{\includegraphics[width=3.2in, height = 1.6in]{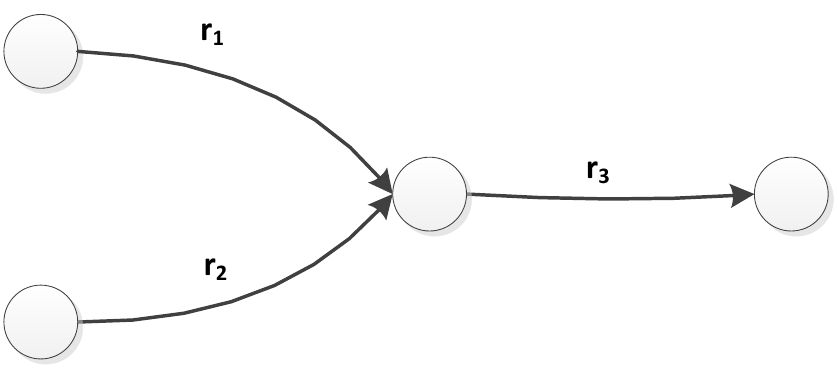}}
			\caption{A schematic diagram of the ``Y'' style traffic network with three road segments.}
			\label{Ynetwork}
		\end{center}
	\end{figure}

	In this section we investigate the impact of pollutants  on pedestrians in different scenarios with networked vehicles. To illustrate the benefits of deploying our extended engine management strategies, namely taking account of both pedestrians data and the traffic flow factor on different roads segments, we consider a symmetric ``Y'' style traffic network with three road segments pictured in Fig. \ref{Ynetwork}. \newline

\begin{figure}[htbp]
		\begin{center}
			{\includegraphics[width=3.2in, height = 2.4in]{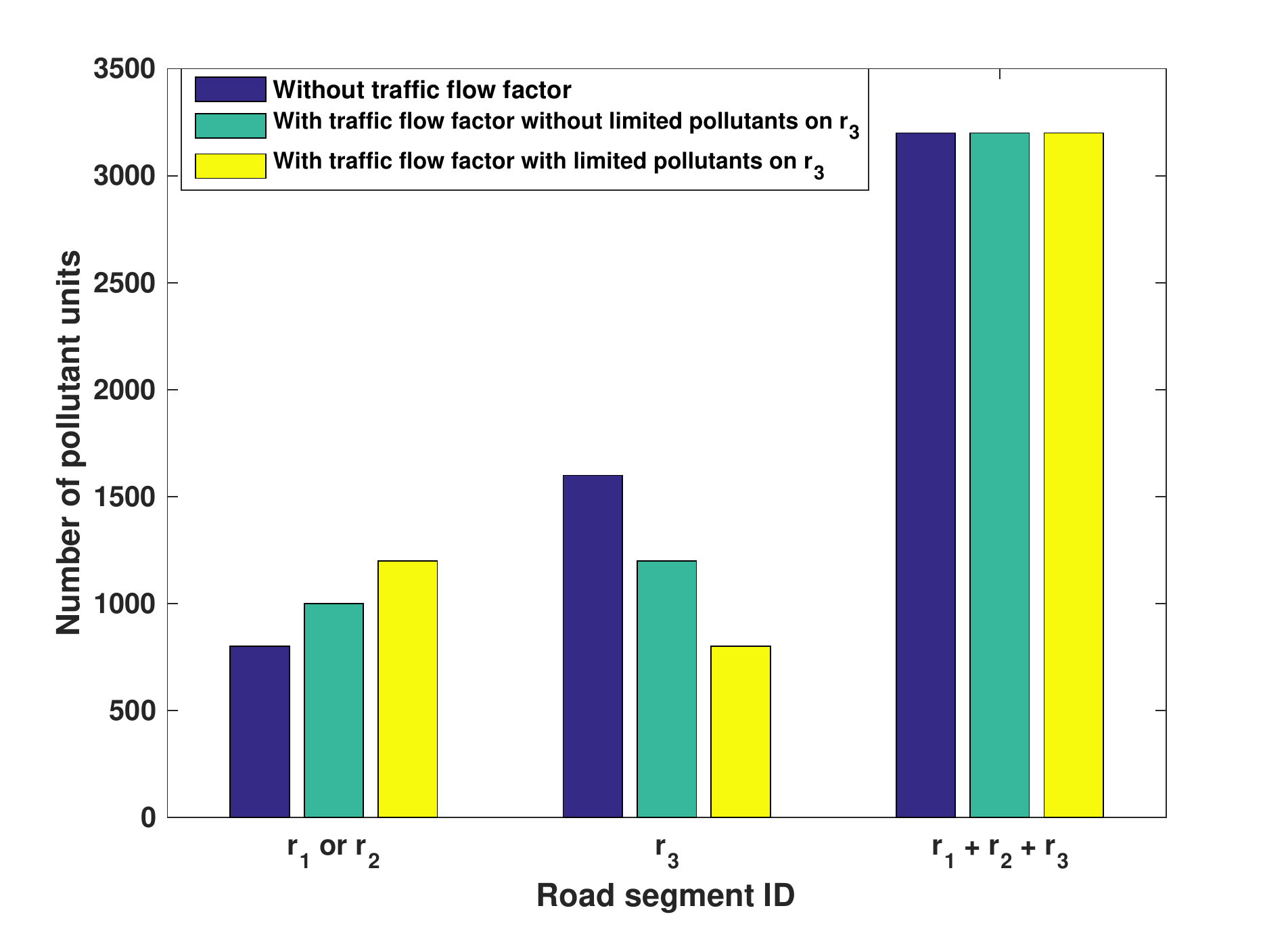}}
			\caption{A comparison diagram for the number of pollutant units with and without considering traffic flow factor.}
			\label{networkResult1}
		\end{center}
	\end{figure}

	In this network, we assume that the vehicles can either start from the beginning of the road segment $r_1$ or $r_2$ but both terminate at the end of the road segment $r_3$, as what is shown in Fig. \ref{Ynetwork}.  We also assume that the average energy consumption on all road segments is the same for all vehicles, which equals $0.025$ kWh, and every PHEV has a same initial energy budget equal to $0.01$ kWh at its own starting point. Most importantly, we assume that the number of pedestrians on each road segment is 50, and the traffic flows on $r_1$, $r_2$ and $r_3$ are assumed to consist of 20, 20 and 40 vehicles at the steady state. In this context, we perform our experimental activities in the following three scenarios: 
	
	\begin{itemize}
		\item[1)] all vehicles solve Problem 1 without considering the traffic flow impact on all road segments;
		\item[2)] all vehicles solve Problem 3 considering the traffic flow factor on all road segments;
		\item[3)] based on 2), a feedback signal is further added to restrict the maximum number of pollutant units on a particular road segment.
	\end{itemize}
	In particular, we call the third scenario ``limited pollutants'' and we assume that the specific road segment in our example is $r_3$ with a maximum number of 800 pollutant units allowed. The comparison results are shown in Fig. \ref{networkResult1}, where we easily observe from the first two scenarios that by simply extending our algorithm by including a traffic flow factor, the number of pollutant units on road segment $r_3$ decreases significantly from 1600 units to 1200 units (25\% less), which tremendously reduces the impact on pedestrians on $r_3$ due to the heavy traffic. Furthermore, by including a feedback signal on $r_3$ in the ``limited pollutants'' scenario, the number of pollutant units further decreases to 800 units, which satisfies all constraints and yields an enhanced performance for our proposed engine management strategy.  \\

	Finally we briefly note that our current framework can also be easily extended to carry out more restricted environmental enforcements ideas in reality, such as the ``green zone'' scenario \cite{naoum2016smart}, in which no pollutant is allowed in a given zone. In this circumstance, the driver can simply reserve the amount of energy required for driving on those segments, and implement the same proposed algorithm to optimally allocate the remaining available energy. To demonstrate this idea, we consider a ``green zone'' scenario in the same network as in Fig. \ref{Ynetwork}. The number of pedestrians and vehicles are assumed to be the same as above. However, we assume that each vehicle has a 0.04 kWh energy budget to be consumed, and specifically, in the ``green zone'' scenario we explicitly require that there should be no pollutants on road segment $r_3$. After the implementation of our algorithm, we obtain that the number of pollutant units on $r_1$, $r_2$, and $r_3$ equals 400, 400, 0, respectively. With a total amount of 800 pollutant units on all road segments, our strategy is very promising to be deployed in practice.

\section{Limitations and extensions} \label{limitations}

The work presented in this paper is an important step towards ``pedestrian-aware engine management". However, it is only a first step and neglects some aspects of a  complete solution, and these will be the subject of future work. For instance, the proposed system does not model the evolution of pollution and realistic dispersion models, as well as the impact of topology. Finally, we used the probability model to account for uncertainty in driver intention. The other uncertainty in the context of electric vehicles is the energy needs of the drivers, and models to capture this uncertainty may be investigated in future work.

\section{Conclusion} \label{conclusion}

In this paper, we have presented a novel engine management system for PHEVs. This management system is designed to benefit those most harmed by automotive emissions; namely,  pedestrians. This is in contrast to traditional algorithms which benefit the polluters; namely, vehicle owners. Also, the proposed approach paves the way to a dynamic and less conservative definition of city restricted areas. We have implemented our proposed strategies in a real test vehicle. Limited field tests have demonstrated the efficacy of the algorithm. The result presented in \cite{guardiola2016adaptive} can be used as a basis for extending the work presented in this paper to ICE vehicles.

\appendix[Markov chain models and driver intention] \label{appendixSection}

Markov chains provide a compact method of storing large volumes of data (assuming certain assumptions are satisfied). Formally speaking, a Markov chain model is a tuple $\langle S, \mathcal{A}, P \rangle$, with $S$ being a finite set of states ($s_1,\ldots, s_n$), $\mathcal{A}$ being a finite set of actions, and $P$ being the transition function $P: S\times \mathcal{A} \times S \rightarrow \mathbb{R}$. In our context, a junction between road segments in a journey defines a node, or state, in a graph $\mathcal{G}$, and the road segment between two nodes defines a link in $\mathcal{G}$. Measured turning probabilities at each junction are used to construct the entries of $P$.  This transition matrix essentially captures journeys from all possible origins to all possible destinations for this driver and can be used to predict driver intention. In particular, the objective of the route prediction engine is then to use this transition matrix $P$ and the vehicle's current position to get all possible road segments to be travelled by the driver and their associated probability values. In this paper we adopted a simpler and equivalent model to the Markov chain approach, to simplify the discussion of the model. However, we recognise that the Markov chain approach would be more efficient in storing details of historical routes of single vehicles. For more details regarding the use of Markov chains for traffic modelling in vehicular applications, see references \cite{SAMITS,krumm2008markov,horvitz2012some,krumm2012people,krumm2013destination}.

\section*{Acknowledgment}

The work of Yingqi Gu, Mingming Liu, and Robert Shorten was supported by Science Foundation Ireland under grant 11/PI/1177.

\bibliographystyle{ieeetran}
\bibliography{References}


\begin{IEEEbiography}[{\includegraphics[width=1in,height=1.25in,clip,keepaspectratio]{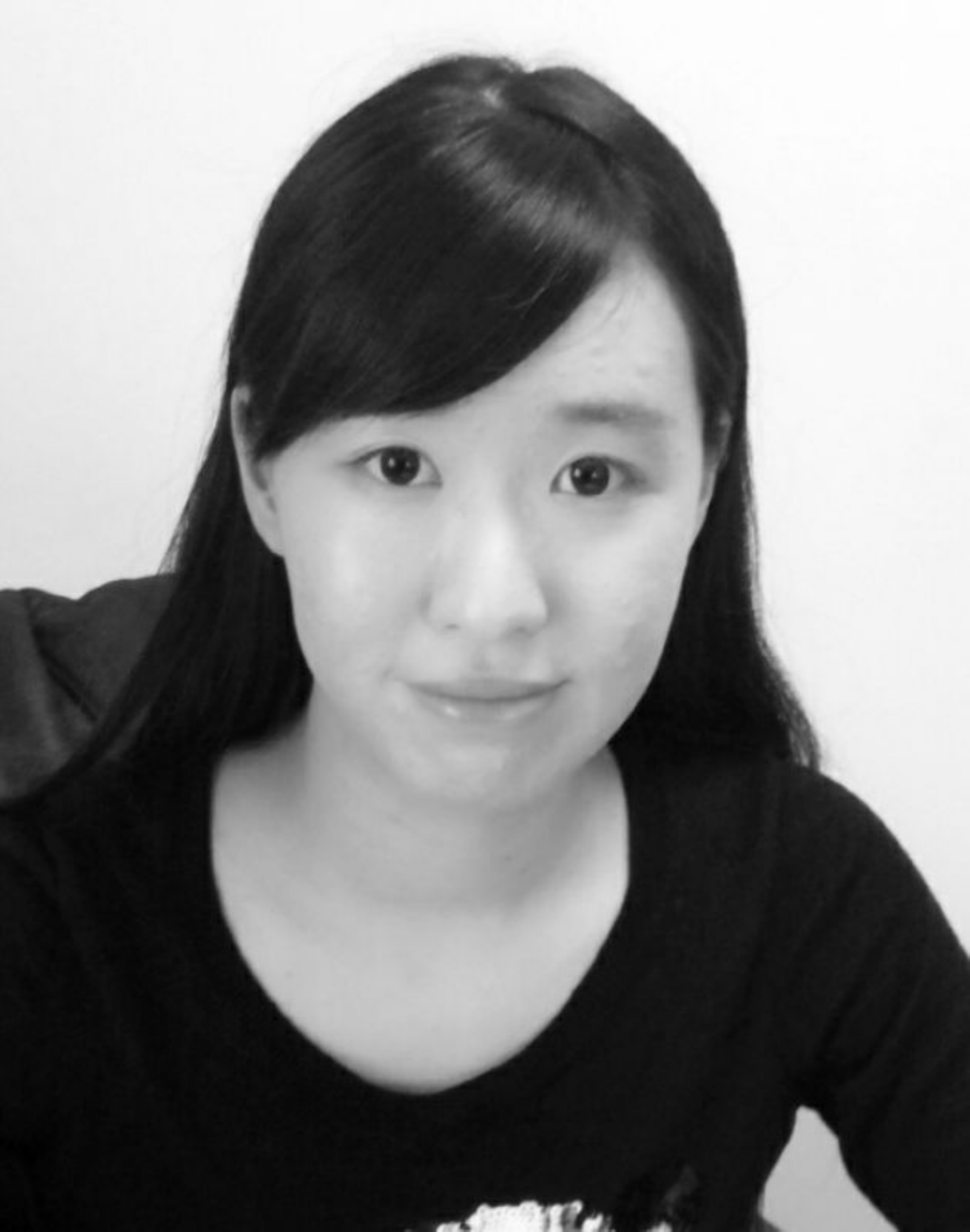}}]{Yingqi Gu}
	received her B.E. degree in Electronic Engineering (first class honours) from Maynooth University in 2013. She obtained her M.Sc degree in Signal Processing and Communications at School of Engineering, University of Edinburgh in 2014. From February 2015, she commenced her Ph.D. degree in the University College Dublin with Prof. Robert Shorten. Her current research interests are modelling, simulation and optimisation of intelligent transportation systems.   		
\end{IEEEbiography}

\begin{IEEEbiography}[{\includegraphics[width=1in,height=1.25in,clip,keepaspectratio]{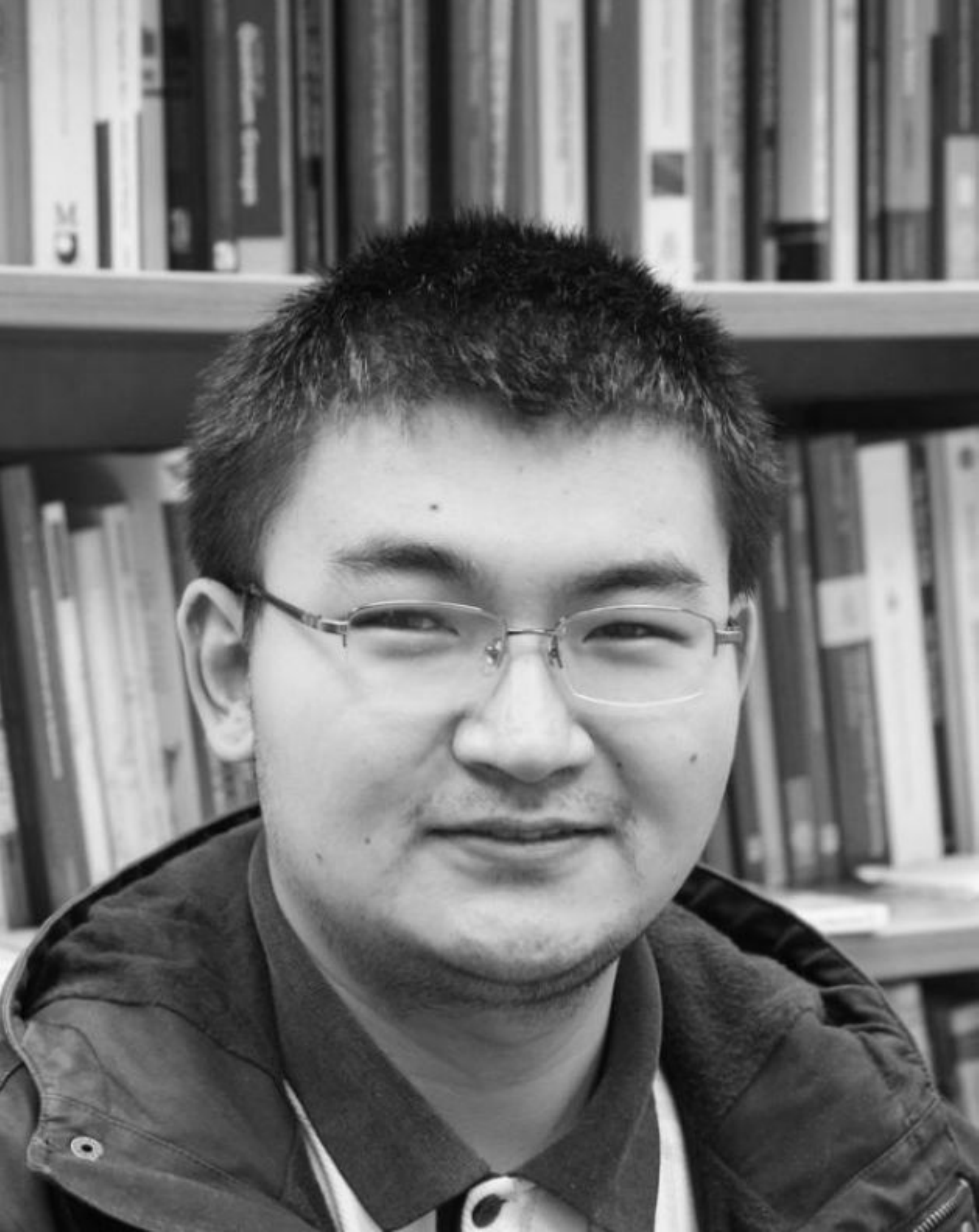}}]{Mingming~Liu}
	received his B.E. degrees in Electronic Engineering with first class honours from National University of Ireland Maynooth in 2011. He obtained his Ph.D. degree from the Hamilton Institute, Maynooth University in 2015. He is currently a post-doctoral research fellow in University College Dublin, working with Prof. Robert Shorten. His current research interests are nonlinear system dynamics, distributed control techniques, modelling and optimisation in the context of smart grid and smart transportation systems. 		
\end{IEEEbiography}

\begin{IEEEbiography}[{\includegraphics[width=1in,height=1.25in,clip,keepaspectratio]{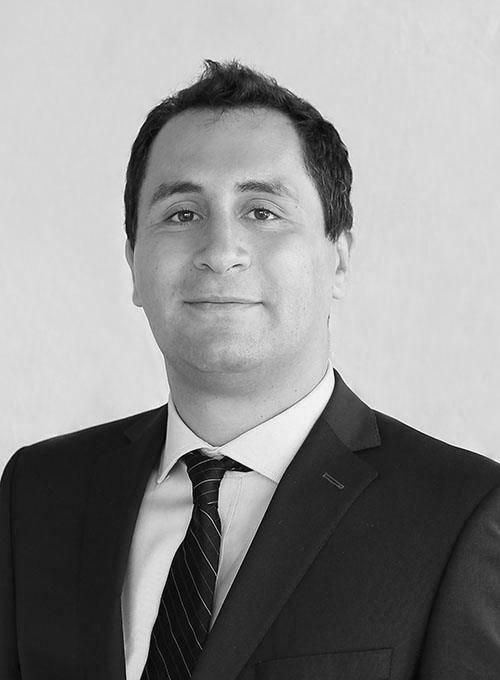}}]{Joe Naoum-Sawaya}
	is currently an Assistant Professor in Management Science at Ivey Business School. He received a B.E. in Computer Engineering from the American University of Beirut and a M.A.Sc. and Ph.D. in Operations Research from the University of Waterloo. His research interests include large scale optimisation methods for practical problems arising in the industry. 
\end{IEEEbiography}

\begin{IEEEbiography}[{\includegraphics[width=1in,height=1.25in,clip,keepaspectratio]{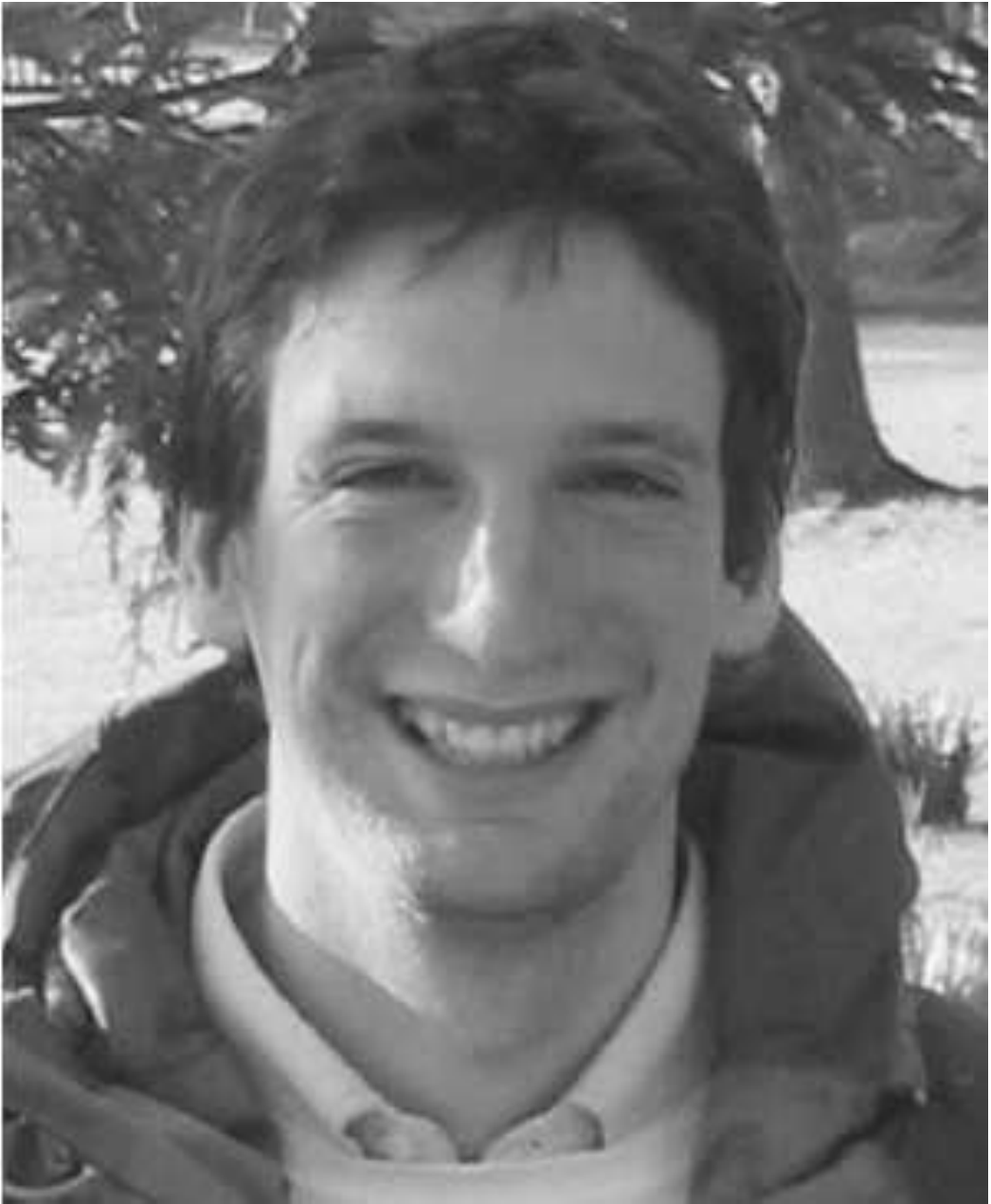}}]{Emanuele Crisostomi}
	received the B.S. degree in computer science engineering, the M.S. degree in automatic control, and the Ph.D. degree in automatics, robotics, and bioengineering, from the University of Pisa, Italy, in 2002, 2005, and 2009, respectively. He is currently an Assistant Professor of electrical engineering with the Department of Energy, Systems, Territory and Constructions Engineering, University of Pisa. His research interests include control and optimisation of large-scale systems, with applications to smart grids and green mobility networks.
\end{IEEEbiography}

\begin{IEEEbiography}[{\includegraphics[width=1in,height=1.25in,clip,keepaspectratio]{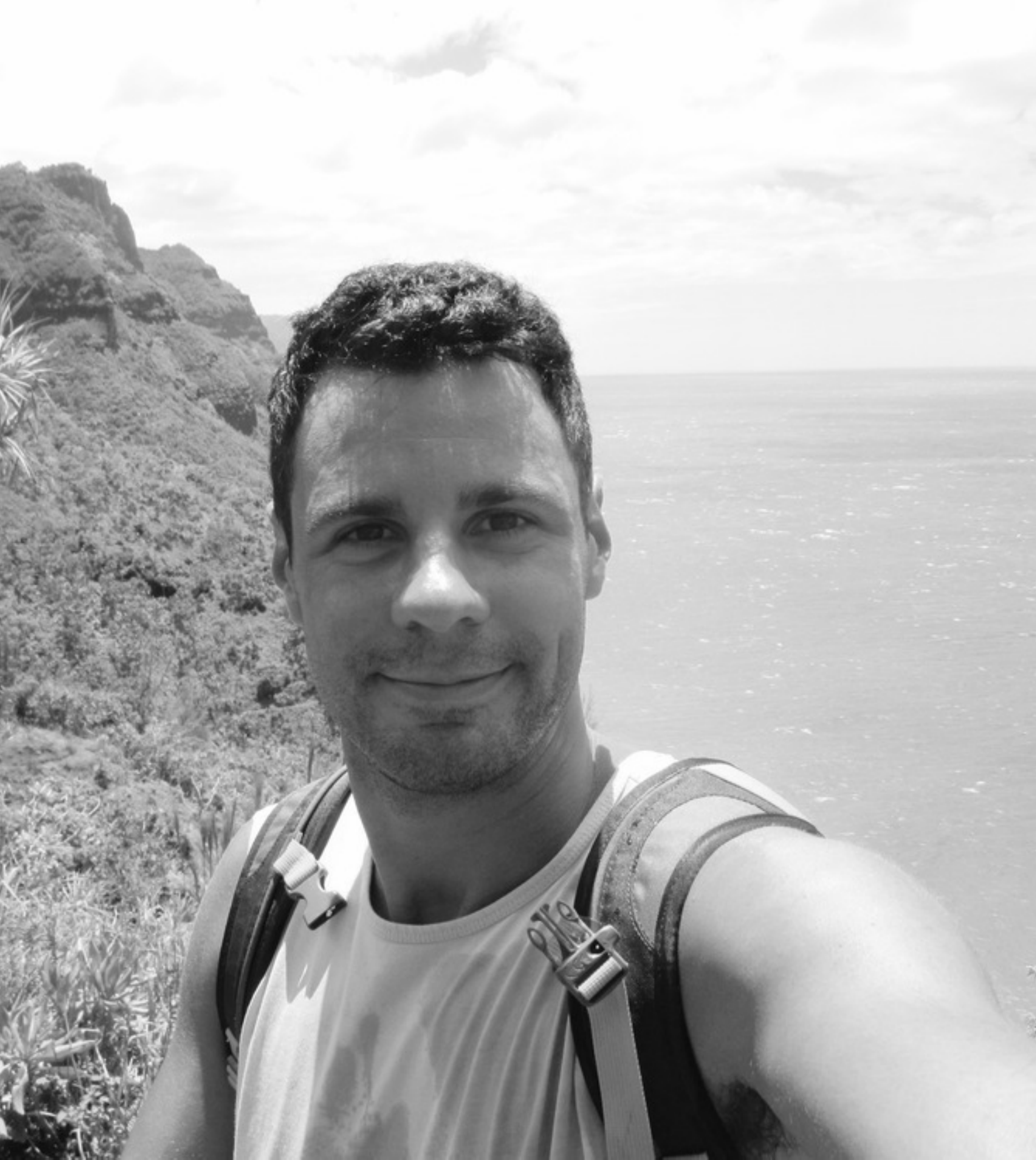}}]{Giovanni Russo}
	
	obtained his Ph.D. from the University
	of Naples Federico II in 2010. During his Ph.D.,
	his work focused on the stability of nonlinear dynamical
	systems with applications to networked control.
	From 2012 to 2015 he has been the lead system engineer and
	integrator of the Honolulu Rail Transit Project (HRTP), the first driverless mass transit railway in the Unites States. In 2015,
	Giovanni joined IBM Research Ireland. Areas of interest include: connected and autonomous
	cars, railways, road traffic control, IoT applications and management of
	shared infrastructures. Giovanni is currently a member of the Board of Editors for the IEEE
	Transactions on Circuits and Systems I for topics related to Control Theory,
	Networks and Networked Systems.
	
\end{IEEEbiography}

\vspace{-15cm}
\begin{IEEEbiography}[{\includegraphics[width=1in,height=1.25in,clip,keepaspectratio]{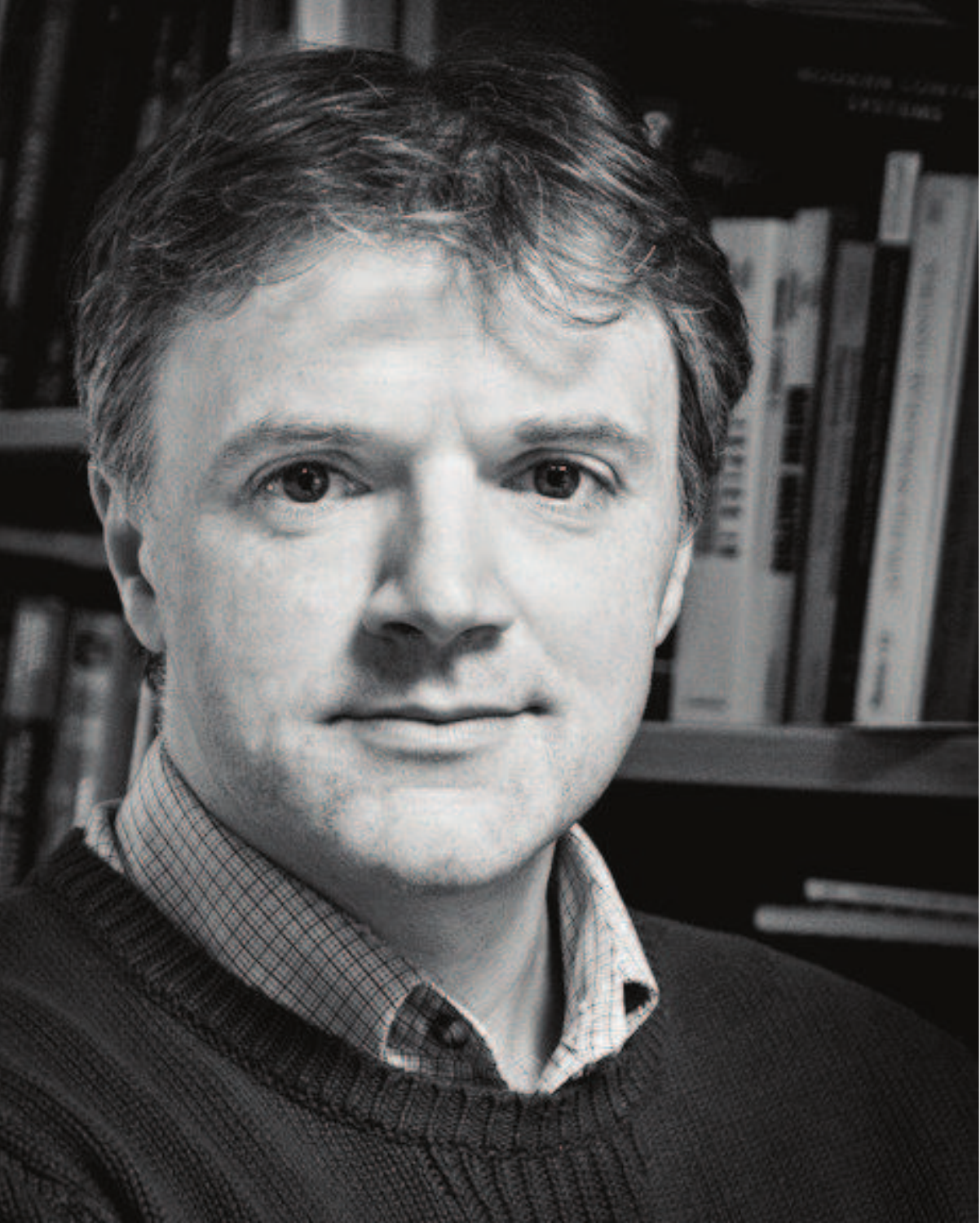}}]{Robert Shorten}
	Professor Shorten is currently Professor of Control Engineering and Decision Science at University College Dublin, and holds a position at IBM Research. Prof. Shorten's research spans a number of areas. He has been active in computer networking, automotive research, collaborative mobility (including smart transportation and electric vehicles), as well as basic control theory and linear algebra. His main field of theoretical research has been the study of hybrid dynamical systems.
\end{IEEEbiography}

\end{document}